\documentclass{article}
\usepackage{alltt}
\usepackage[all]{xy}
\usepackage{latexsym,amssymb,amsmath,amsfonts, amsthm, xcolor}

\def\vbar{\mathchoice{\vrule height6.3ptdepth-.5ptwidth.8pt\kern-.8pt}
  {\vrule height6.3ptdepth-.5ptwidth.8pt\kern-.8pt}
  {\vrule height4.1ptdepth-.35ptwidth.6pt\kern-.6pt}
  {\vrule height3.1ptdepth-.25ptwidth.5pt\kern-.5pt}}
\def\fudge{\mathchoice{}{}{\mkern.5mu}{\mkern.8mu}}
\def\bbc#1#2{{\rm \mkern#2mu\vbar\mkern-#2mu#1}}
\def\bbb#1{{\rm I\mkern-3.5mu #1}}
\def\bba#1#2{{\rm #1\mkern-#2mu\fudge #1}}
\def\bb#1{{\count4=`#1 \advance\count4by-64 \ifcase\count4\or\bba A{11.5}\or
  \bbb B\or\bbc C{5}\or\bbb D\or\bbb E\or\bbb F \or\bbc G{5}\or\bbb H\or
  \bbb I\or\bbc J{3}\or\bbb K\or\bbb L \or\bbb M\or\bbb N\or\bbc O{5} \or
  \bbb P\or\bbc Q{5}\or\bbb R\or\bbc S{4.2}\or\bba T{10.5}\or\bbc U{5}\or
  \bba V{12}\or\bba W{16.5}\or\bba X{11}\or\bba Y{11.7}\or\bba Z{7.5}\fi}}

\newcommand{\vs}{\vspace{0.25cm}}

\newtheorem{theorem}{Theorem}
\newtheorem{itlemma}{Lemma}[section]
\newtheorem{itproposition}[itlemma]{Proposition}
\newtheorem{itcorollary}[itlemma]{Corollary}
\newtheorem{itremark}[itlemma]{Remark}
\newtheorem{itremarks}[itlemma]{Remarks}
\newtheorem{itdefinition}[itlemma]{Definition}
\newtheorem{itexample}[itlemma]{Example}

\newenvironment{lemma}{\begin{itlemma}\rm}{\end{itlemma}} 
\newenvironment{remark}{\begin{itremark}\rm}{\end{itremark}} 
\newenvironment{remarks}{\begin{itremarks} \rm}{\end{itremarks}}
\newenvironment{corollary}{\begin{itcorollary}\rm}{\end{itcorollary}}
\newenvironment{proposition}{\begin{itproposition}\rm}{\end{itproposition}}
\newenvironment{definition}{\begin{itdefinition}\rm}{\end{itdefinition}}
\newenvironment{example}{\begin{itexample}\rm}{\end{itexample}}
\newenvironment{fact}{\noindent {{\bf Fact}}:\ \ }{\hfill \medskip}
\newenvironment{claim}{\noindent {\em Claim}. \ \ }{\hfill \medskip}
\newcommand{\be}[1]{\begin{equation}\label{#1}}
\newcommand{\ee}{\end{equation}}
\newcommand{\bl}[1]{\begin{lemma}\label{#1}}
\newcommand{\br}[1]{\begin{remark}\label{#1}}
\newcommand{\brs}[1]{\begin{remarks}\label{#1}}
\newcommand{\bt}[1]{\begin{theorem}\label{#1}}
\newcommand{\bd}[1]{\begin{definition}\label{#1}}
\newcommand{\bp}[1]{\begin{proposition}\label{#1}}
\newcommand{\bc}[1]{\begin{corollary}\label{#1}}
\newcommand{\bfact}[1]{\begin{fact}\label{#1}}
\newcommand{\bex}[1]{\begin{example}\label{#1}}
\newcommand{\ec}{\end{corollary}}
\newcommand{\efact}{\end{fact}}
\newcommand{\eex}{\end{example}}
\newcommand{\el}{\end{lemma}}
\newcommand{\er}{\end{remark}}
\newcommand{\ers}{\end{remarks}}
\newcommand{\et}{\end{theorem}}
\newcommand{\ed}{\end{definition}}
\newcommand{\ep}{\end{proposition}}
\newcommand{\epr}{\end{proof}}
\newcommand{\bpr}{\begin{proof}}
\newcommand{\bcl}{\begin{claim}}
\newcommand{\ecl}{\end{claim}}

\newcommand{\bi}{\begin{itemize}}
\newcommand{\ei}{\end{itemize}}
\newcommand{\ben}{\begin{enumerate}}
\newcommand{\een}{\end{enumerate}}

\usepackage{graphicx}

\title{\bf \Large{Sub-Riemannian geodesics on $SU(n)/S(U(n-1) \times U(1))$ and optimal control of three level quantum systems}
}
\vs

\vs

\author{Francesca Albertini,\thanks{Dipartimento di Analisi e Gestione dei Sistemi Industriali,  Universit\`a di Padova, albertin@math.unipd.it}  \, \, Domenico D'Alessandro,\thanks{Department of Mathematics, Iowa State University, Ames, Iowa, U.S.A., e-mail:daless@iastate.edu} \, \, Benjamin Sheller \thanks{Department of Mathematics, Iowa State University, Ames, Iowa, U.S.A., e-mail:bsheller@iastate.edu}}

\begin{document}

\maketitle

\begin{abstract}

We study the time optimal control problem for the evolution operator 
of an $n$-level quantum system from the identity to any desired final condition.  For the  considered class of quantum systems the control couples all the energy levels to a given one and is assumed  to be bounded in Euclidean  norm. From a mathematical perspective,  such a problem is a sub-Riemannian $K-P$ problem, as introduced in \cite{Ugo}, \cite{JurdjeBasic}, whose underlying symmetric space is  
$SU(n)/S(U(n-1) \times U(1))$. Following the method of \cite{NOIJDCS},  we consider the action of $S(U(n-1) \times U(1))$ on $SU(n)$ as a conjugation $X \rightarrow AXA^{-1}$. This  allows us to do a symmetry 
reduction and consider the problem on a quotient space. We give an explicit description of such a quotient space which has the structure of a stratified space. We prove several properties of sub-Riemannian problems with the given structure. We derive the explicit optimal control for the case of three level quantum systems where the desired operation is on the lowest two energy levels ($\Lambda$-systems). We solve this latter problem by reducing it to an integer quadratic  optimization problem with linear constraints.   

\end{abstract}

\section{Introduction}

Many finite dimensional quantum systems of interest 
in applications can be modeled by the {\it Schr\"odinger operator equation} in the form \cite{Mikobook}
\be{SCRODOP}
\dot X =A X +\sum_{j=1}^m B_j X u_j, \qquad X(0)={\bf 1}, 
\ee 
where $u_j$ are {\it control electromagnetic semi-classical fields} 
which can be decided by an experimenter. The unitary matrix $X$ is the {\it evolution operator} (or {\it propagator}) of the quantum mechanical systems and $A$ and $B_j$ are matrices in the Lie algebra $\mathfrak{su}(n)$ where $n$ is the number of energy levels of the system. 
Typically, one  works in the basis of the eigenvectors of the `{\it internal Hamiltonian}' $A$, so that $A$ is diagonal, while the $B_j$'s represent couplings between different levels activated by the external fields $u_j$. Such couplings are described by the {\it energy level diagram} of the system (see, e.g., \cite{Sakurai}).  In (\ref{SCRODOP}), the matrix ${\bf 1}$ represents the identity. We shall assume a special structure for systems (\ref{SCRODOP}) in the following, which, among other things, guarantees the {\it controllability} of the system, that is, every special unitary matrix can be reached from the identity in finite time. We shall be interested in finding the optimal field control $u_j$'s, bounded as 
\be{sumsquare}
\sum_{j=1}^m u_j^2 \leq \gamma^2, 
\ee  
which drives $X$ from the identity ${\bf 1}$ to a desired final condition $X_f \in SU(n)$ in {\it minimum time}. Consideration of minimum time control is natural in quantum mechanics applications to computation in that  one would like to perform computational tasks as quickly as possible. Additionally, and more in general, fast evolution is a way to counteract the negative effect of the environment (de-coherence) in quantum dynamics \cite{Breuer}, so as to fully exploit the quantum behavior of a given system.

\begin{figure}[htb]
\centering
\includegraphics[width=0.65\textwidth, height=0.35\textheight]{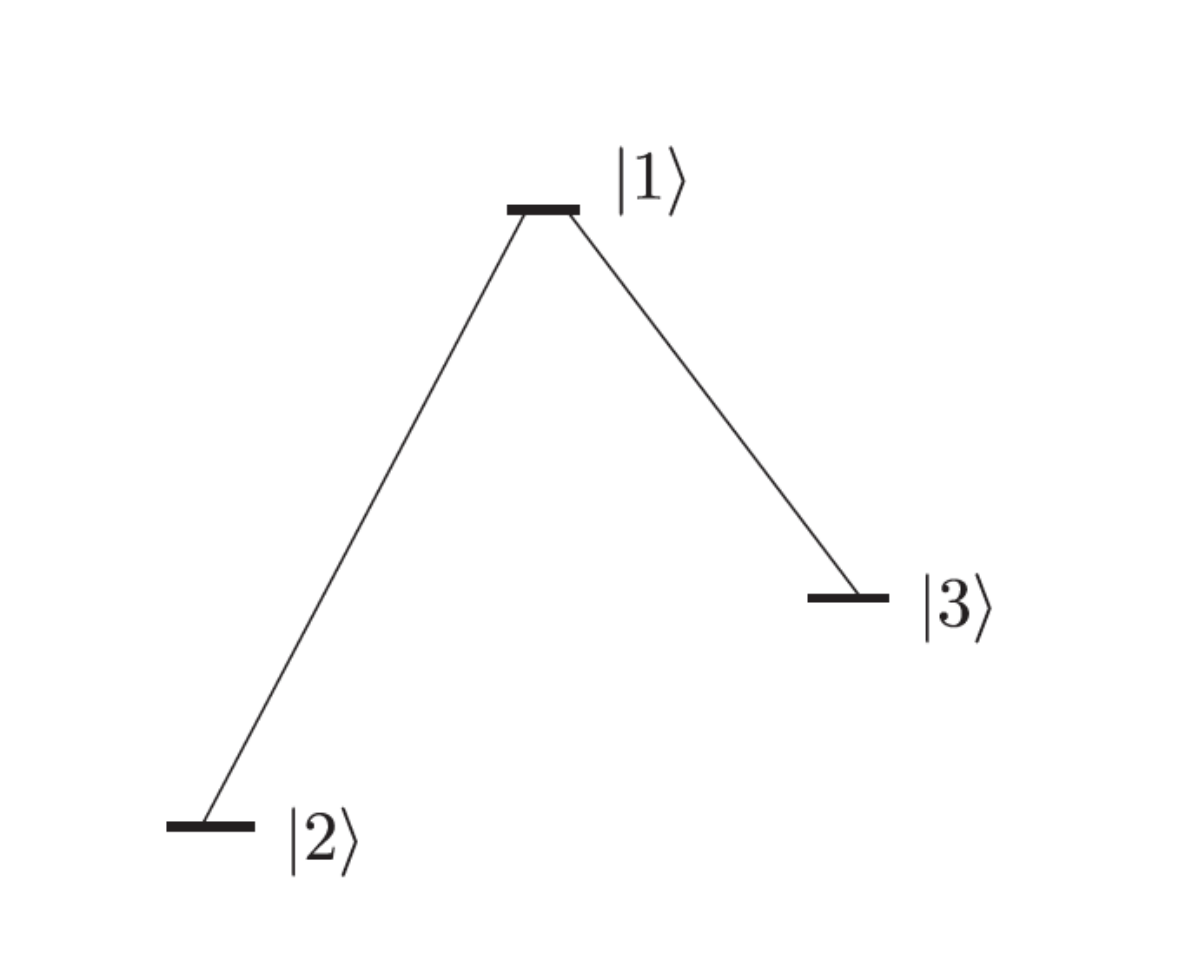}
\caption{Energy level diagram for a $\Lambda$-system}.
\label{Fig0}
\end{figure}

In this paper, we shall consider a special class of systems whose  energy level diagram  couples one energy level to all the others. An example of this, which will be treated in detail, is the class of so-called $\Lambda$-systems where the highest energy level is coupled to the lowest two levels (cf. Figure \ref{Fig0}) but the lowest  energy levels are not coupled with each other directly. {$\Lambda$} systems, \cite{French} \cite{Eberly}, 
arise in many contexts, including molecular dynamics and quantum optics as well as systems exhibiting electromagnetically induced transparency \cite{Reed} .  

For the class of $n-$level quantum systems in (\ref{SCRODOP}) we shall consider,  $A$ is a diagonal matrix, while the $B_j$'s are matrices, modeling the couplings, that are all zero except in the $(1,l)$-th entry for $l=2,3,\ldots,n$ (and corresponding $(l,1)$-th entry) which are given by $\frac{1}{\sqrt{2}}$ (and $-\frac{1}{\sqrt{2}}$) or  $\frac{i}{\sqrt{2}}$ (and $\frac{i}{\sqrt{2}}$). In particular we have  $m=2(n-1)$ in (\ref{SCRODOP}). With this choice,  the $B_j$'s are orthonormal with respect to the inner product on $\mathfrak{su}(n)$, $\langle C,D \rangle :=Tr(CD^\dagger)$, and they are orthogonal to the diagonal matrix $A$ which is also assumed normalized with respect to the norm corresponding to this inner product. Normalization of $A$ and the $B_j$'s can be obtained by re-scaling the time and-or the controls in the problem.  

By going to the {\it interaction picture} \cite{Sakurai}, i.e., defining the new propagator $U:=e^{-At}X$ we can effectively eliminate the drift in equation (\ref{SCRODOP}). The equation for $U$ is 
\be{neweq}
\dot U=\sum_{j=1}^m e^{-At}B_je^{At} u_j U, \qquad U(0)={\bf 1}.
\ee 
As a consequence of the structure of equation (\ref{SCRODOP}), for each $j$,
 $e^{-At}B_je^{At}$ is a linear combination of the $B_j$'s, i.e., 
\be{lincombi}
e^{-At}B_je^{At}:=\sum_{k=1}^m a_{j,k}(t)B_k, 
\ee
and replacing this into (\ref{neweq}) and defining the new controls 
$v_k:=\sum_{j=1}^m a_{j,k} u_j$ the equation becomes the {\it driftless} equation 
\be{neweq2}
\dot U=\sum_{k=1}^m B_k v_k U, \qquad U(0):={\bf 1}.
\ee
Notice that this does not change the nature of the minimum time problem since $\| v\|=\|u\|$.

Our main concern in this paper will be the minimum time problem for system (\ref{neweq2}). This problem is related to the corresponding problem for the system with a nonzero drift (\ref{SCRODOP}) as follows. First, if the task is to obtain an operation which achieves the transfer between two eigenvectors of the matrix $A$, the minimum time obtained for the 
system (\ref{neweq2}) will be the same as the  minimum for the system (\ref{SCRODOP}) since if $\psi_U$ is the state for the system (\ref{neweq2}), ($\psi_U(t)=U(t)\psi_U(0)$) the state for system (\ref{SCRODOP}) will be $\psi(t)=X(t)\psi(0)=e^{At} \psi_U(t)$. This differs from $\psi_U(t)$ only by a phase factor and therefore is still an eigenvector of $A$. Moreover, even for general final conditions $X_f$ in $SU(n)$, the complete optimal synthesis for system (\ref{neweq2}) 
gives the complete optimal synthesis for (\ref{SCRODOP}). In fact, for a fixed $t$, the set of states reachable in minimum time $t$ for system (\ref{neweq2}) gives the boundary of the reachable set ${\cal R}_U(t)$  for system (\ref{neweq2}). The reachable set for system (\ref{SCRODOP}) at time $t$ is ${\cal R}(t)=e^{At} {\cal R}_U(t)$. Knowledge of the reachable set at any time is equivalent to the knowledge of the minimum time for any desired final condition.

The minimum time problem for system (\ref{neweq2}) with bounded $L_2$ norm of the control is equivalent to the problem, for fixed $T$, to minimize the `energy' $\int_0^T\|v(t)\|^2 dt$ . The problem is also equivalent to finding the sub-Riemannian geodesics on $SU(n)$ where the sub-Riemannian structure is specified by the vector fields $B_j$'s (see, e.g., \cite{ABB}, \cite{Agrachev}, \cite{NOIJDCS}). The optimal control with the bound $\| v \|^2 \leq 
\gamma^2$ is the same as the optimal control with bound   
\be{newbound2}
\| v \|^2 \leq 
2
\ee
multiplied by $\frac{\gamma}{\sqrt{2}}$ and time scaled by $\frac{\gamma}{\sqrt{2}}$ and therefore the bound in (\ref{sumsquare}) can always be normalized. We shall assume therefore in the following the bound (\ref{newbound2}). Furthermore the optimal control is such that equality always holds in (\ref{sumsquare}) (cf., e.g., \cite{NOIJDCS}).

In the paper \cite{JurdjeBasic} V. Jurdjevi\'c introduced a class of problems for which the problem on $SU(n)$ above described is a special case. Consider a semisimple Lie algebra ${\cal L}$ which has a (Cartan)  decomposition ${\cal L}:={\cal K} \oplus {\cal P}$, satisfying the commutation relations 
\be{COMM}
[{\cal K}, {\cal K}] \subseteq {\cal K}, \qquad [{\cal K}, {\cal P}] \subseteq {\cal P}, \qquad 
 [{\cal P}, {\cal P}] \subseteq {\cal K}.  
\ee  
Then the minimum time problem for the system (\ref{neweq2}), where the $B_j$'s form an orthonormal basis in ${\cal P}$, is called a {\it $K-P$ problem}. Such terminology was in fact used in \cite{Ugo} where the application to quantum systems was first considered. The problem for quantum systems considered in this paper  is a $K-P$ problem with the Lie algebra ${\cal L}$ given by $\mathfrak{su}(n)$, the subalgebra ${\cal K}$ given by block diagonal matrices in $\mathfrak{su}(n)$ with blocks of dimension $1$ and $n-1$, and the complementary subspace ${\cal P}$ spanned  by anti-diagonal matrices with the corresponding dimensions of the blocks. Several other possibilities may occur and in fact E. Cartan has classified, up to change of coordinates, all the possible Cartan decompositions for the classical groups \cite{Helgason}.  In \cite{JurdjeBasic} V. Jurdjevi\'c also gave the form of the optimal control and trajectories for $K-P$ problems. By applying the Pontryagin Maximum Principle in the version for systems on Lie groups (see, e.g, \cite{Sachkov}) one finds that there exist a matrix $P \in {\cal P}$ and a matrix $A \in {\cal K}$ such that the optimal control $v_j$ in (\ref{neweq2}) satisfies 
\be{2Bsat}
\sum_{k=1}^m B_k v_k(t)=e^{At} P e^{-At}, 
\ee 
from  which one obtains the components $v_k$ of the optimal control, assuming that  the matrices $B_k$ give  an orthonormal basis of ${\cal P}$. Moreover by solving the equation $\dot U=(\sum_{k=1}^m B_k v_k(t))U=e^{At} P e^{-At}U,$ $U(0)={\bf 1}$ the corresponding optimal trajectory is given by 
\be{Optimaltraj}
U(t)=e^{At}e^{(-A+P)t}. 
\ee 
The problem of finding the optimal control to reach a final state $X_f$ becomes therefore 
the problem of finding matrices $A\in{\cal K}$ and $P \in {\cal P}$ and real $ t> 0$ such that $X_f:=e^{At}e^{(-A+P)t}$ and $t$ is the minimum positive value such that this is possible. In the $SU(n)$ case, this involves the  search for  $n^2-1$ parameters: the $n^2-1$ parameters in the matrices $A$ and $P$, plus the parameter $t$, minus the normalization of $P$ due to the fact that $\| P\|=\|v\|=\gamma$ is fixed.


The objective of this paper is to solve the above optimal control problem for a three level quantum system, and in fact we will provide explicit  optimal control laws for a set of final conditions of interest in applications. In this setting this study is an extension of the results provided in \cite{Ugo}, to more general final conditions and with a different method.  In the process,  we shall prove several general properties for the optimal control for system (\ref{neweq2}) for general problems on $SU(n)$. We follow the approach in \cite{NOIJDCS} in that we consider the action of the Lie group $S(U(n-1) \times U(1))$, block diagonal matrices in $SU(n)$ with blocks of dimension $1$ and $n-1$, on $SU(n)$ by conjugation. This allows us to reduce the optimal control problem to the orbit  space of $SU(n)/S(U(n-1) \times U(1))$ and, for practical purposes, to reduce the number of parameters to search in the matrices $A$ and $P$ in (\ref{Optimaltraj}) which can in fact be assumed to be of a special form (Proposition \ref{tridiag}). Every optimal trajectory can be found as a `lift' to $SU(n)$ of a geodesic  in the orbit space  $SU(n)/S(U(n-1) \times U(1))$.  In section \ref{App1} we prove some general properties of the given optimal control problem and in section \ref{App2} we give a characterization of the orbit space   $SU(n)/S(U(n-1) \times U(1))$ where the projections of geodesics are to be found.  Starting from  section \ref{App3} we focus on  the case of a {\it three} level quantum $\Lambda$-system and on final conditions which are operators on the Hilbert subspace corresponding to the lowest two energy levels. In this section we reduce the problem of optimal control to a quadratic  integer optimization problem with linear constraints. We solve such a problem in section \ref{solupro}. In section \ref{concluese} we give an example and a discussion of the results.

{

\section{Symmetry reduction for the K-P problem $SU(n)/S(U(n-1)\times U(1))$}\label{App1}
Consider first a general $K-P$ problem as described in the introduction (cf. (\ref{neweq2}) and (\ref{COMM})) and the action of the Lie group associated with\footnote{We follow the convention of denoting by $e^{\cal K}$ the connected (component containing the identity of the) Lie group associated with the Lie algebra ${\cal K}$.} ${\cal K}$, $e^{\cal K}$, on the Lie group associated with ${\cal L}$, $e^{\cal L}$, by left (or, equivalently, right) multiplication. Then the quotient space $e^{\cal L}/e^{\cal K}$ is a {\it symmetric space} \cite{Helgason} for $e^{\cal L}$ and this was the original motivation for V. Jurdjevi\'c to introduce such problems in \cite{JurdjeBasic}. The sub-Riemannian structure on ${\cal L}$ induces a Riemannian structure and metric on   $e^{\cal L}/e^{\cal K}$. Riemannian geodesics on  $e^{\cal L}/e^{\cal K}$ lift to sub-Riemannian (SR) geodesics on $e^{\cal L}$ (cf. Lemma 26.11 in \cite{Michor1}). For  the systems considered here, the Lie group $e^{\cal K}$ is the Lie group of block diagonal matrices with blocks of dimension $1$ and $n-1$ and determinant equal to $1$, $S(U(n-1)\times U(1))$. The quotient space space $SU(n)/S(U(n-1)\times U(1))$ is isomorphic to the complex projective space $\mathbb{CP}^{n-1}$ which, in quantum mechanics,  represents the space of pure states. The induced Riemannian metric on $\mathbb{CP}^{n-1}$ is the Fubini-Study metric \cite{Nomizu} (Chapter XI, Example 10.5), and the Riemannian geodesics on $\mathbb{CP}^{n-1}$ are known (cf. \cite{Nomizu} pg. 277).\footnote{The Riemannian geometry on  $\mathbb{CP}^{n-1}$ is the main tool for modern derivations of `quantum speed limits' and `energy-time uncertainty relations' in the physics literature (cf. \cite{Brazil} and the references therein).} Therefore SR geodesics in $SU(n)$ can be found by parallel lifting of the known geodesics on  
$\mathbb{CP}^{n-1} \equiv SU(n)/S(U(n-1)\times U(1))$. However, only a special class of  SR geodesics on $SU(n)$ can be found in this manner, these are sometimes called `parallel geodesics'. Since we are interested in the {\it full} optimal control problem on $SU(n)$,  we shall explore a different type of reduction  obtained by considering a different action of $S(U(n-1) \times U(1))$ on $SU(n)$, that is,  {\it conjugation}. An element $K \in S(U(n-1) \times U(1))$ acts on an element $X$ in  $SU(n)$ by $X \rightarrow K X K^{-1}=KXK^\dagger$. If we consider this 
action, {\it all} geodesics on $SU(n)$ are projected to 
geodesics on $SU(n)/S(U(n-1)\times U(1))$, which is however 
no longer a manifold but rather has the more general structure of a stratified space \cite{Bredon}. Riemannian geometry on singular spaces was studied 
in \cite{Dimitry}. As proved in \cite{NOIJDCS}, the complete optimal synthesis for the problem on the original manifold (in this case $SU(n)$) can be obtained by analyzing the  geodesics on the quotient space associated to the problem (in this case $SU(n)/S(U(n-1)\times U(1))$). In fact, if $U_d:=U_d(t)$ is a minimum time trajectory for (\ref{neweq2}) with final condition $U_f$, then $KU_dK^{-1}:=KU_dK^{-1}(t)$ is the minimum time trajectory (with the same time) with final condition $KU_fK^{-1}$. By an analysis on the quotient space, one can obtain information about optimal geodesics, reachable sets, and 
various geometrical loci of interest. These include the {\em{cut locus}}, locus of points reached by two or more geodesics, and the {\em{critical locus}}, locus of points where geodesics lose optimality. 
This approach was followed in \cite{Noibas} to solve the optimal control problem for the case of two level quantum systems. In that case the quotient space was  (homeomorphic to) the closed unit disc in the complex plane and the generalization of this fact  for the case $n \geq 3$ will be presented in Section \ref{App2}. 

For general $K-P$ problems, the action of $e^{\cal K}$ on $e^{\cal L}$ by conjugation  lifts to an action (which is also conjugation) of $e^{\cal K}$ on the Lie algebra $\mathcal{L}$, which, when restricted to ${\cal K}$ is the adjoint representation of $e^{\cal K}$ on the $\cal K$ part. The subspace $\cal P\subseteq {\cal L} $ is also invariant under conjugation by $e^{\cal K}$ because of the second formula in (\ref{COMM}). This symmetry reduction allows for a reduction of the number of parameters to be determined to find the optimal control law, i.e., the parameters  in the matrices $A$ and $P$ in (\ref{2Bsat}) (\ref{Optimaltraj}). This is because  we only need to consider a single representative in the equivalence class of any geodesic. By multiplying (\ref{Optimaltraj}) on the left and right by $K$ and $K^{-1}$ respectively, we see that the matrices $A$ and $P$ can be chosen up to a common conjugation by an element $K \in e^{\cal K}$. Therefore there is no loss of generality to consider $A$ and $P$ of a special form (see Proposition \ref{tridiag}} below). In particular, by fixing  a maximal Abelian subalgebra $\cal{A}$ in $\cal P$ (a Cartan subalgebra), we can assume that   $P$ is an element of $\cal{ A}$. This is because $\cal P$ may be written as $\mathcal P=\cup_{K\in e^{\cal K}} K{\cal{A}} K^{-1}$ (c.f. Proposition 7.29 in \cite{Knapp 1}). In the case of $SU(n)$ with $\cal P$ the orthogonal complement of $\mathcal K=\{\text{block diagonal matrices with first block } 1\times 1\text{ other block } (n-1)\times (n-1)\}$, we have that any Cartan  subalgebra ${\cal A}$ is one dimensional. Therefore, it suffices to fix one non-zero matrix $P\in\cal P$ and only consider the control problem with that particular $P$. We remark in particular that we must have  $\| P\|^2=2$ because of (\ref{2Bsat}) and (\ref{newbound2}).  We will then take $P$ to be the matrix with $(1,2)$ entry equal to $i$ (and hence $(2,1)$ entry 
also equal to $i$) and all other entries equal to zero. We can also assume a special, tridiagonal,  form for $A \in {\cal K}$ in (\ref{2Bsat}), (\ref{Optimaltraj}) as described in the following Proposition. In the following we shall use ${\cal K}$ to denote the Lie algebra of block diagonal matrices in $\frak{su}(n)$ with blocks of dimension $1$ and $n-1$ and ${\cal P}$ the complementary subspace of antidiagonal matrices.

\bp{tridiag}
Let $A\in\cal{K}$ and $P\in\cal{P}$ with $P$ having $(1,2)$ and $(2,1)$ entries nonzero and all others zero. Then $-A+P$ may be tridiagonalized by a  special unitary matrix with a $2\times 2$ identity matrix in the upper left corner, which is therefore in particular a matrix in $e^{\cal K}=S(U(n-1) \times U(1))$. Furthermore, the off-diagonal entries of the tridiagonalized form, starting from row $2$ may be taken purely  imaginary (or purely real) and are nonzero if $A$ has a nonzero nondiagonal entry in the corresponding row.

\ep


\bpr
The proof proceeds by induction on the size $n\times n$ of the matrices. First, note that for $2\times 2$ matrices, the result holds as the matrix $-A+P$ is already tridiagonal, and we have assumed (without loss of generality) $P$ purely imaginary. Now, suppose that $n\geq 3$. Let $-A+P=(b_{ij})_{1\leq i,j\leq n}$ and let $\hat v:=0$ if $b_{32}=...=b_{n2}=0$ and $\hat v:=i(\sum_{k=3}^n |b_{k2}|^2)^{-1}$ otherwise. Let $S=(s_{ij})_{1\leq i,j\leq n-2}\in SU(n-2)$ such that $s_{1j}=\frac{\bar{b}_{j+2,2}}{\hat{v}^*}$ if $\hat v\neq 0$ and $S\in SU(n-2)$ arbitrary otherwise. Then $S^{\dagger}V=B$ where $V$ is the vector with first entry equal to $\hat v$ and all other entries zero and $B$ the vector with entries $(b_{j2})_{3\leq j\leq n}$. Therefore, the following equation holds:
\[
\left(\begin{array}{lc}
I_2 & 0 \\
0 & S \\
\end{array}\right)
\left(\begin{array}{ccccc}
b_{11} & b_{12} & 0 & \cdots & 0 \\
b_{21} & b_{22} & &  \cdots & b_{2n} \\
0 & \vdots & & &\vdots \\
\vdots & & & & \\
0 &b_{n2} & & \cdots & b_{nn}\\
\end{array}\right)
\left(\begin{array}{lc}
I_2 & 0 \\
0 & S^{\dagger}\\
\end{array}\right)=
\left(\begin{array}{cccccc}
b_{11} & b_{12} & 0 &\cdots  & & 0\\
b_{21} & b_{22} & -\hat{v}^*& 0 & \cdots & 0\\
0 & \hat{v} & * &\cdots & * & *\\
0 & 0 &* &\cdots & * & *\\
\vdots & \vdots & \vdots& & &\vdots \\
0 & 0 & *&\cdots & &*\\
0 & 0 &*&\cdots&&*\\
\end{array}\right)
\]
Considering now the submatrix
\[
\left(\begin{array}{ccccc}
b_{22} & -\hat{v}^* & 0 &\cdots & 0\\
\hat v & * &* & \cdots & *\\
0 &* & * &\cdots& *\\
\vdots & \vdots &&&\vdots\\
0 & *&\cdots &&*\\
0 & *&\cdots &&*\\
\end{array}\right)
\]
and recursively using the result with $n$ replaced by $n-1$ proves the claim. The last claim of the proposition follows from the choice of $\hat v$ at each step. 
\epr
From the proof of the proposition and in particular the nature of $\hat v$ as a purely imaginary number we have the following. 
\bc{Noloss}
There is no loss of generality in assuming that $-A+P$ is a purely imaginary tridiagonal 
matrix in $su(n)$ and therefore so are $A \in {\cal K}$ and $P \in {\cal P}$.     
\ec
This in particular means that we can always write $-A+P$ as $i$ multiplied by a real symmetric tridiagonal matrix.

\vs

In the following (sections \ref{App3} and \ref{solupro}), we shall focus  on optimal control problems for final conditions 
in $e^{\cal K}$.  These final conditions correspond to operations on the lowest $n-1$ levels  of the given quantum systems. In this case (and for general $K-P$ problems) in order for $e^{At}e^{(-A+P)t}$ in (\ref{Optimaltraj}) to lie in $e^{\cal K}$, multiplying on the left by $e^{-At}$ yields that $e^{(-A+P)t}\in e^{\cal K}$. The following property is instrumental in finding the optimal control. 

\vs

\bp{scalar}
Consider the K-P decomposition of $SU(n)$ as above. Suppose $A\in\cal{K}$ and $P\in\cal{P}$ such that $-A+P$ is a tridiagonal matrix with no elements of the sub-(or super-)diagonal equal to zero, and suppose $e^{(-A+P)t}\in e^{\cal K}$. Then $e^{(-A+P)t}$ is a scalar matrix.
\ep 
\bpr
We proceed by induction on $n$. First, observe that the result holds for $n=2$ by directly computing the matrix exponential:
\[
\exp\left(\begin{array}{lc}
	ait  &  (c+di)t \\
	(-c+di)t &  bit \\
	  \end{array}\right)=
\frac{e^{it(a+b)/2}}{\omega}\left(\begin{array}{lc}
	\omega\cos(t\omega/2)+(a-b)i\sin(t\omega/2) &  2(c+id)\sin(t\omega/2) \\
	2(-c+id)\sin(t\omega/2) &  \omega\cos(t\omega/2)-(a-b)i\sin(t\omega/2)\\
	  \end{array}\right)
	  \]
where $\omega=\sqrt{(a-b)^2+c^2+d^2}$ and $c+di\neq 0$. If this exponential has off-diagonal entries equal to zero, then $t\omega/2=k\pi$ for some $k\in\mathbb{Z}$, so the only possibility for the exponential is a matrix with $\pm e^{it(a+b)/2}$ on the diagonal (with the same sign) and zeros elsewhere. Now, consider $n>2$.
Let $U=e^{(-A+P)t}$. Then $-A+P$ commutes with $U$. Observe that $e^{\cal K}$ acting on $\frak{su}(n)$  by conjugation fixes both $\cal K$ and $\cal P$ (from \ref{COMM}).  It also therefore fixes the natural extensions of these subspaces to subspaces of $\mathfrak{u}(n)$. Since  $[U,-A+P]=0$ we have $[U,A]=[U,P]$, which is equivalent to $UAU^\dagger-A=UPU^\dagger-P$. Since the left hand side 
of this equality is in ${\cal K}$ and the right hand side is in ${\cal P}$ then both sides are zero and therefore $[U,A]=[U,P]=0$. Now, using the special form of $P$ (which is zero everywhere except in the $(1,2)$ and $(2,1)$ entry which are equal to $i$) we have that $[U,P]=0$ implies $(U)_{2,2}=(U)_{1,1}$ and $(U)_{2,k}=0$ for $k>2$. So $U$ is not only in $e^{\cal K}$ but it has a block diagonal form with a   $2 \times 2$ (upper left) block which is a scalar matrix. Now decompose:
\[
A:=A_n=\left(\begin{array}{lc}
	a_{11}  &  0     \cdots   0 \\
	0 &   \\
	 \vdots &\hat{A}_{n-1} \\
	  0 & \\ \end{array}\right),
	  \]
where $\hat{A}_{n-1}=A_{n-1}+P_{n-1}$ with
\[
A_{n-1}=\left(\begin{array}{lc}
	a_{22}  &  0     \cdots   0 \\
	0 &   \\
	 \vdots &\hat{A}_{n-2} \\
	  0 & \\ \end{array}\right),\hspace{1cm}
P_{n-1}=\left(\begin{array}{ccc}
	0  &  a_{23}   & 0 \cdots   0 \\
	-\bar{a}_{23} &  & \\
	0 &  &\\
	 \vdots & &\text{\huge0} \\
	  0 & &\\ \end{array}\right)
	  \]
Since $U$ commutes with $A$, the lower $n-1 \times n-1$ block of $U$, $U_{n-1}$, commutes with $\hat A_{n-1}=A_{n-1}+P_{n-1}$. Proceeding as above, by replacing $U$ with $U_{n-1}$ 
and using the fact that $a_{23}$ is  different from zero we find that $(U)_{3,3}=(U)_{2,2}$ and $(U)_{3,k}=0$ for $k >3$. Proceeding in this fashion inductively we find that $U$ is a scalar matrix. 
\epr
\bc{scalcor}
If $A\in\cal{K}$ with each row other than the first one having at least one non-zero entry on the off-diagonal, and $P\neq 0$, then $e^{(-A+P)t}\in e^{\cal K}$ for some $t\in\mathbb{R}$ if and only if $e^{(-A+P)t}$ is a scalar matrix.
\ec
\bpr
Propositions \ref{tridiag} and \ref{scalar} above show that $e^{(-A+P)t}$ must be similar to a scalar matrix; but any scalar matrix is in the center, so $e^{(-A+P)t}$ must actually be a scalar matrix.
\epr

\section{The orbit space $SU(n)/S(U(n-1)\times U(1))$}\label{App2}
Since optimal trajectories are the inverse images (via the natural projection $\pi: SU(n) \rightarrow SU(n)/S(U(n-1)\times U(1))$ of optimal trajectories on the quotient space, it is important to understand the geometric nature of such a space. Then we can study and visualize  optimal geodesics on this space each  corresponding to a family of optimal geodesics $\{ KU(t)K^\dagger | K \in e^{\cal K}\}$ in $SU(n)$.  According to general results in the theory of transformation groups \cite{Bredon} $SU(n)/S(U(n-1)\times U(1))$ has the structure of a stratified space. In this section we study the structure of such a space.

The Lie group $S(U(n-1)\times U(1))$ can be parametrized as:
\be{equiv0}
\left( \begin{array}{lc}  
		e^{i\eta} & 0 \\
		0 &\xi V  \end{array} \right),
\ee
with $V \in SU(n-1)$, $\eta \in [0,2\pi)$ and $\xi\in\mathbb{C}$ such that $\xi^{n-1}=e^{-i\eta}$. To simplify the notation, we denote by 
\[
SU(n)/_{\sim}=SU(n)/S(U(n-1)\times U(1))
\]
the orbit space, so two matrices in $ SU(n)$ are in 
the same orbit if they are conjugated by an element of $S(U(n-1)\times U(1))$. 
To describe the orbit space $SU(n)/_{\sim}$ we also need to 
consider the following equivalence relation in $SU(n)$:
\be{equiv1}
X_1\sim_{\phi} X_2 \  \   \  \Leftrightarrow       \        \    \ \exists\, U\in SU(n) \text{ such that } \  \ 
U\left( \begin{array}{lc}  
		e^{i\phi} & 0 \\
		0 & I \end{array} \right) X_1U^\dagger = \left( \begin{array}{lc}  
		e^{i\phi} & 0 \\
		0 & I \end{array} \right) X_2,  
\ee
where $I$ denotes here the $(n-1) \times (n-1)$ identity. Equivalently for a fixed $\phi$, 
$X_1$ and $X_2$ are $\sim_\phi$ equivalent if and only if $\left( \begin{array}{lc}  
		e^{i\phi} & 0 \\
		0 & I \end{array} \right) X_1$ and $\left( \begin{array}{lc}  
		e^{i\phi} & 0 \\
		0 & I \end{array} \right) X_2$ have the same spectrum. We define a  
		topological fiber bundle on the circle $S^1$ as
$\pi:E_n\to S^1 $ with fibers $\pi^{-1}(e^{i\phi})=SU(n-1)/_{\sim_{\phi}}$.
The fibers $\pi^{-1}(e^{i\phi})=SU(n-1)/_{\sim_{\phi}}$ 
may not be manifolds, but are merely topological spaces with 
the coarsest topology that makes $\pi$ a continuous map. An example is given in Proposition \ref{Mobius} for the case $n=3$ where each fiber is a closed segment $[-1,1]$, a {\it manifold with boundary}. 

Let $D_1$ be the open unit disc in $\mathbb{C}$, i.e. if  $x\in D_1$ then $x$ is  a complex number with absolute value strictly less than 1. The next theorem recursively describes  the orbit space
$SU(n)/_{\sim}$. 
\bt{equiv2}
Let $\Psi$ be the map from $E_n\cup \left(D_1\times SU(n-1)/_{\sim}\right)$ to  $SU(n)/_{\sim}$, defined by:
\be{equiv3a}
\text{ if } [Z]_{\sim_{\phi}}\in E_n \  \ \text{ then } \  \ \Psi\left([Z]_{\sim_{\phi}}\right)=
\left[ \left(\begin{array}{lllll} 
	e^{-i\phi} &  0 & 0& \cdots &0 \\
	0  & e^{i\phi} & 0& \cdots &0 \\
	0 &  0 & &&\\
	\vdots &\vdots& & I &\\
	0& 0  & & & 
	 \end{array} \right) \left(\begin{array}{lc}
	1 &  0     \cdots   0 \\
	0 &   \\
	 \vdots & Z \\
	  0 & \\ \end{array}\right) \right]_{\sim}
\ee
\[
\text{ if } \left( x, [Z]_{\sim} \right) \in D_1\times SU(n-1)/_{\sim} \  \  \text{ then }  \  \  \hspace{6.5cm}
\]
\be{equiv3b}
\Psi \left( x, [Z]_{\sim} \right) = 
\left[ \left(\begin{array}{cclll} 
	x &  \sqrt{1-|x|^2} & 0& \cdots &0 \\
	-\sqrt{1-|x|^2}   & x^* & 0& \cdots &0 \\
	0 &  0 & &&\\
	\vdots &\vdots& & I &\\
	0& 0  & & & 
	 \end{array} \right) \left(\begin{array}{lc}
	1 &  0     \cdots   0 \\
	0 &   \\
	 \vdots & Z \\
	  0 & \\ \end{array}\right) \right]_{\sim}.
\ee
Here, we are gluing the fiber $\pi^{-1}(e^{i\varphi})$ in $E_n$ to the point $e^{i\varphi}$ in the boundary of $D_1$. Then the map $\Psi$ is a global homeomorphism.
\et

 If we identify $SU(1)/_{\sim}$ and   $SU(1)/_{\sim_\phi}$ with a single point so 
 that $E_2\equiv S^1=\partial D_1$, we have that (with $\simeq$ denoting homeomorphism) $SU(2)/_{\sim}\simeq  \partial D_1 \cup D_1=\bar D_1$ is homeomorphic to the  {\it closed} unit disc. 
This  result  was known from the case of the optimal control of two level quantum systems \cite{Noibas} and  Theorem \ref{equiv2} is a generalization to the $n-$level case. 
 Applying Theorem \ref{equiv2} recursively, we obtain. 
\bc{expli1} With $E_1$ equal by definition to a single point, we have, for $n \geq 2$ 
\be{rhs123}
SU(n)/_{\sim}\simeq \bigcup_{j=0}^{n-1} D_1^{\times j} \times E_{n-j}.
\ee
\ec
The number of parameters characterizing the equivalence class depends on the subset in the right hand side of (\ref{rhs123}) where the equivalence class is. In particular, 
if $\Psi^{-1}([X]_{\sim}) \in  D_1^{\times j} \times E_{n-j}$, for $j=0,...,n-1$, the number of parameters is $n+j-1=(n-j-1)+2j$, with $n-j-1$ the number of parameters for $E_{n-j}$ and $2j$ the number of parameters for $\times^j D_1$.

 
\vs



\bpr (Proof of Theorem \ref{equiv2}) 

We need to prove that  $\Psi$ is i) well-defined, ii) onto, iii) one-to-one, and iv) continuous with continuous inverse.
Here we only prove that  $\Psi$  is onto, since its argument shows how to find $\Psi$-inverse and therefore how to parametrize the elements in $SU(n)/\sim$. We postpone to the Appendix the remaining  part of the proof.

{\bf {$\Psi$ is onto.}}

We need to show that for any $X_f\in SU(n)$ there exists an element $Y\in E_n\cup \left(D_1\times SU(n-1)/_{\sim}\right)$ such that $\Psi(Y)=[X_f]_{\sim}.$ Write $X_f$ using the Cartan  decomposition of type {\bf AIII} of $SU(n)$ \cite{Helgason} (i.e., (\ref{COMM}) where ${\cal K}$ are block diagonal matrices in $\frak{su}(n)$ with blocks of dimensions $1$ and $n-1$) as:
 \be{equiv6}
 X_f=K_1MK_2,
 \ee
 where 
 \be{equiv7}
 M=
\left(\begin{array}{ccccc} 
	\cos(\theta)& \sin(\theta) & 0& \cdots &0 \\
	-\sin(\theta)  & \cos(\theta)& 0& \cdots &0 \\
	0 &  0 & &&\\
	\vdots &\vdots& & I &\\
	0& 0  & & & 
	 \end{array} \right) 
\ee	 
and
\be{equiv8}
K_j=\left( \begin{array}{lc}  
		e^{i\eta_j} & 0 \\
		0 & e^{-\frac{i\eta_j}{n-1}}V_j  \end{array} \right),  \  \   \  j=1,2, \  \  V_j\in SU(n-1).
\ee
\noindent {\em{Claim}}: Without loss of generality we can 
assume $\sin(\theta)   \geq 0$.
\noindent {\em{Proof of the Claim}}: Assume  $\sin(\theta)<0$, and let
\[
L=\left(\begin{array}{ccccc} 
	i& 0 & 0& \cdots &0 \\
	0  & -i& 0& \cdots &0 \\
	0 &  0 & &&\\
	\vdots &\vdots& & I &\\
	0& 0  & & & 
	 \end{array} \right).
	 \]
Then $X_f= {K}_1LL^\dagger{M}LL^\dagger K_2$. Letting $\hat{K}_1={K}_1L$,
$\hat{K}_2=L^\dagger K_2$, and $\hat{M}=L^\dagger{M}L$, we have that 
$X_f=\hat{K}_1\hat{M}\hat{K}_2$, $\hat{K}_j\in S(U(n-1)\times U(1))$, and 
$\hat{M}$ is equal to ${M}$ except for  the sign of  the off-diagonal elements. So the Claim is proved. Given 
$X_f=K_1MK_2,$ where $M$ is as in 
(\ref{equiv7}), with $\sin(\theta) \geq 0$, and $K_i \in S(U(n-1)\times U(1))$, we have:
\[
[X_f]_{\sim}=[K_1^\dagger K_1 M K_2 K_{1}]_{\sim}=[M{K_2 K_1}]_{\sim}.
\]
Therefore,  to conclude  that $\Psi$ is onto, it is enough to prove that for every element of the type;
\be{equiv10}
\left(\begin{array}{ccccc} 
	\cos(\theta)& \sin(\theta) & 0& \cdots &0 \\
	-\sin(\theta)  & \cos(\theta)& 0& \cdots &0 \\
	0 &  0 & &&\\
	\vdots &\vdots& & I &\\
	0& 0  & & & 
	 \end{array} \right) 
\left( \begin{array}{lc}  
		e^{i\phi} & 0 \\
		0 & e^{-\frac{i\phi}{n-1}}V  \end{array} \right),
		\ee
\noindent with $\sin(\theta)\geq 0$, there exists an $\sim$-equivalent element of the form
\be{equiv10bis}
 \left(\begin{array}{cclll} 
	x &  \sqrt{1-|x|^2} & 0& \cdots &0 \\
	-\sqrt{1-|x|^2}   & x^* & 0& \cdots &0 \\
	0 &  0 & &&\\
	\vdots &\vdots& & I &\\
	0& 0  & & & 
	 \end{array} \right) 	\left(\begin{array}{lc}
	1 &  0     \cdots   0 \\
	0 &   \\
	 \vdots &Z \\
	  0 & \\ \end{array}\right),
	 \ee
with $Z\in SU(n-1)$ and $|x|\leq 1$. Here, if $|x|<1$ the pre-image of the equivalence class of the above  matrix will be  in 
$\left(D_1\times SU(n-1)/_{\sim}\right)$, while, if $|x|=1$, it  will be in $E_n$.

Consider the matrix:
\[
F:=\left(\begin{array}{ccccc} 
	e^{i\frac{\phi}{2}}& 0 & 0& \cdots &0 \\
	0 & e^{-i\frac{\phi}{2}}& 0& \cdots &0 \\
	0 &  0 & &&\\
	\vdots &\vdots& & I &\\
	0& 0  & & & 
	 \end{array} \right),
	 \]
then the following matrix is equivalent to (\ref{equiv10}):
\be{equiv11a}
F \left(\begin{array}{ccccc} 
	\cos(\theta)& \sin(\theta) & 0& \cdots &0 \\
	-\sin(\theta)  & \cos(\theta)& 0& \cdots &0 \\
	0 &  0 & &&\\
	\vdots &\vdots& & I &\\
	0& 0  & & & 
	 \end{array} \right) F F^{\dagger}
\left( \begin{array}{lc}  
		e^{i\phi} & 0 \\
		0 & e^{-\frac{i\phi}{n-1}}V  \end{array} \right)F^{\dagger}.
		\ee
We have:
\be{equiv11}
F \left(\begin{array}{ccccc} 
	\cos(\theta)& \sin(\theta) & 0& \cdots &0 \\
	-\sin(\theta)  & \cos(\theta)& 0& \cdots &0 \\
	0 &  0 & &&\\
	\vdots &\vdots& & I &\\
	0& 0  & & & 
	 \end{array} \right) F=\left(\begin{array}{ccccc} 
	e^{i\phi}\cos(\theta)& \sin(\theta) & 0& \cdots &0 \\
	-\sin(\theta)  & e^{-i\phi}\cos(\theta)& 0& \cdots &0 \\
	0 &  0 & &&\\
	\vdots &\vdots& & I &\\
	0& 0  & & & 
	 \end{array} \right),
	 \ee
which, by setting $x=e^{i\phi}\cos(\theta)$, is of the same form as the first matrix in equation
(\ref{equiv10bis}), since $0\leq \sin(\theta)=\sqrt{1-|x|^2}$.
Moreover, it follows:
\be{equiv11b}
 F^{\dagger}
\left( \begin{array}{lc}  
		e^{i\phi} & 0 \\
		0 & e^{-\frac{i\phi}{n-1}}V  \end{array} \right)F^{\dagger}=
	\left(\begin{array}{lc}
	1 &  0     \cdots   0 \\
	0 &   \\
	 \vdots &Z \\
	  0 & \\ \end{array}\right),
	  \ee
which is of the same type as the second matrix in equation (\ref{equiv10bis}).

	\epr
The proof of Theorem \ref{equiv2} and formulas (\ref{equiv3a}) and (\ref{equiv3b}) suggest how to invert the homeomorphism $\Psi$ and to find the parameters characterizing the equivalence class of a given matrix $X \in SU(n)$. Given a 
matrix $X \in SU(n)$, one writes its Cartan decomposition, see equation (\ref{equiv6}),  $X=K_1M K_2$,  with $M$ and $K_i$ as in (\ref{equiv7})-(\ref{equiv8}), and 
 with  $\sin(\theta) \geq 0$. Then 
 $[X]_{\sim}=[MK_2 K_1]_{\sim}$. If $|\cos(\theta)|=1$ then the equivalence class of the 
matrix $X$ is in the image of $E_n$, in  particular $[X]_{\sim}= \Psi\left([Z]_{\sim_{\phi}}\right)$  where $\phi= \eta_2+\eta_1$ (mod $\pi$), with $\eta_1$ and $\eta_2$ defined in (\ref{equiv8}), and $Z$ is defined as in equation 
(\ref{equiv11b}).
If $|\cos(\theta)|<1$ then the equivalence class of the matrix is in the image of $D_1 \times SU(n-1)/_{\sim}$, in particular $[X]_{\sim}= \Psi\left(x,[Z]_{\sim}\right)$ where 
$x=e^{i\phi}\cos \theta$ (see (\ref{equiv11})), and $Z$  is defined as before by  equation 
(\ref{equiv11b}).

In the following, we shall focus on the case $n=3$, which concerns the  three level quantum system case.  Applying  the above theorem, we get that  $SU(3)/_{\sim}$ is  characterized by at most 4 parameters. Geometrically we can visualize $SU(3)/_{\sim} \equiv E_3 \cup (D_1 \times SU(2)/_{\sim})$ as follows. Consider a closed disc $\bar D_1$.  At each interior point of the disc, we have another  closed unit  disc representing $SU(2)/_{\sim}$, so at most $4$  parameters. The boundary of the disc serves as the base for the topological fiber bundle $E_3$. Each fiber of $E_3$ is given by a segment $[-1,1]$. The following proposition describes $E_3$ in more detail. 
 
\bp{Mobius} The topological fiber bundle $E_3$ is the closure of the M\"obius band.  
\ep 
\bpr Consider the fiber $\pi^{-1}(e^{i \phi})$ which is the set of equivalence classes of elements  $X$ in $SU(2)$ such that $\begin{pmatrix} e^{i\phi} & 0 \cr 0 & 1 \end{pmatrix} X$ has a given spectrum. Since all elements   $\begin{pmatrix} e^{i\phi} & 0 \cr 0 & 1 \end{pmatrix} X$ have the same determinant $e^{i\phi}$, the equivalence class of $X$ is uniquely determined by the trace  
 $Tr \left(\begin{pmatrix} e^{i\phi} & 0 \cr 0 & 1 \end{pmatrix} X \right)$ or, equivalently, by 
  $\frac{e^{-i\frac{\phi}{2}}}{2}Tr \left(\begin{pmatrix} e^{i\phi} & 0 \cr 0 & 1 \end{pmatrix} X \right)$. By writing $X$ as $X:=\begin{pmatrix}re^{i\psi} & y \cr -y^* & re^{-i\psi} \end{pmatrix}$, we have 
\[ 
\frac{e^{-i\frac{\phi}{2}}}{2}Tr \left(\begin{pmatrix} e^{i\phi} & 0 \cr 0 & 1 \end{pmatrix} X \right)=r\cos(\frac{\phi}{2}+ \psi). 
\]  
As this value varies in $[-1,1]$ as $r$ and $\psi$ change, each fiber is identified with the interval $[-1,1]$. Therefore $E_3$ is parametrized by $s:=r\cos(\frac{\phi}{2}+ \psi) \in [-1,1]$ and $\phi\in [0, 2\pi]$, i.e., a rectangle with $s_0=r\cos(\psi)$ and $s_{2\pi}=r \cos (\psi+\pi)=-r \cos(\psi)=-s_0$ identified. That is, $E_3$ is a M\"obius strip.\footnote{Notice that the curve $(\phi, s(\phi))=(\phi, r \cos(\frac{\phi}{2}+\psi))$ for fixed $r\not=0 $ and $\psi$, as $\phi$ goes from $0$ to $2\pi$,  crosses the line $s=0$ only once. Therefore there is only one `twist' of the rectangle $ 0 \leq \phi \leq 2\pi$, $-1\leq s \leq 1$ before joining the two ends.}  

\epr

\section{Optimal synthesis for three level quantum systems as an integer optimization problem}\label{App3}

 From this point on, we shall restrict ourselves to three level quantum systems in the $\Lambda$ configuration of Figure \ref{Fig111} assuming that we want to perform a desired {\it operation} on the subspace spanned by the lowest two energy eigenstates $|2\rangle$, $|3 \rangle$. In this setting we extend the work of \cite{Ugo} which dealt with the problem of transferring between the two eigenstates $|2\rangle$ and $|3 \rangle$. Our problem is at the level of the evolution operator (this is called the problem `upstairs' in \cite{Ugo}) and for general operations, although later we shall restrict ourselves to matrices with a special structure in order to reduce the number of subcases. The desired final condition is therefore of the form 
\be{Xfdes}
X_{f}:=\begin{pmatrix} * & 0 \cr 
0 & \tilde X_f,  \end{pmatrix}
\ee
with $\tilde X_f \in U(2)$ the desired final transformation on the subspace corresponding to the lowest two energy levels. Because of the symmetry described in the previous sections it is enough to drive the state $U$ in (\ref{neweq2}) time optimally to a matrix {\it similar} to $\tilde X_f$. Once this problem has been solved, by an appropriate similarity transformation we obtain the trajectory and control for the desired final condition.  Therefore the problem is characterized by assigning the two eigenvalues of   $\tilde X_f$ which we denote by $e^{i\alpha}$ and $e^{i\beta}$ with $\alpha$ and $\beta$ in $(-\pi, \pi]$. After symmetry reduction, the problem is (cf. section \ref{App1}) to find real values $a$, $b$, and $c$ and minimum  $t>0$ such that, with 
\be{AandP}
A:=\begin{pmatrix} i(a+b) & 0 & 0 \cr 
0 & -ia & -ic \cr 
0 & -ic & -ib  \end{pmatrix}, \qquad P:=\begin{pmatrix} 0 & i & 0 \cr 
i & 0 & 0 \cr 
0 & 0 & 0 \end{pmatrix}, 
\ee
$e^{At}e^{(-A+P)t}$ in the same equivalence class as (\ref{Xfdes}).  
Notice that since $X_f \in e^{\cal K}=S(U(n-1) \times U(1))$ according to Proposition \ref{scalar}, except for the case where the final condition is in the same class as $diag(e^{i\varphi},e^{-i\varphi},1)$  (in these cases $c=0$), $e^{(-A+P)t}$ is necessarily a scalar matrix. We shall exclude final conditions in the same class as $diag(e^{i\varphi},e^{-i\varphi},1)$. Therefore neither $\alpha$ nor $\beta$, above is zero and we shall use the fact   that $e^{(-A+P)t}$ is a scalar matrix. Furthermore,  since a common phase factor is physically irrelevant in quantum mechanics, we shall  consider the problem to assign eigenvalues $e^{i\alpha}$ and $e^{i\beta}$ to $e^{At}$, in minimum time while ensuring that the matrix $e^{(-A+P)t}$ is scalar.

\subsection{A nonlinear integer optimization problem}

\vs

Since $e^{(-A+P)t}$ must be a scalar matrix  the matrix $(-A+P)t$ must have eigenvalues $\lambda_1$, $\lambda_2$ and $\lambda_3$ of the form  
\be{L1L2L3}
\lambda_1:=i\frac{2k\pi}{3}, \quad \lambda_2:=i\frac{2k\pi}{3}+i2m\pi, 
\quad \lambda_3:=-i\frac{4k\pi}{3}-i2m\pi 
\ee
for integers $k$ and $m$, where we used the fact that $0=\lambda_1+\lambda_2+\lambda_3$. This is true if and only if the symmetric real  matrix 
\be{tildeA}
-i(-A+P):=\begin{pmatrix} -(a+b) & 1 & 0 \cr 1& a & c \cr 0 & c & b\end{pmatrix}, 
\ee 
has eigenvalues in 
\be{eigen}
-i \frac{\lambda_1}{t}=\frac{2k\pi}{3t}, \quad 
-i \frac{\lambda_2}{t}=\frac{2k\pi}{3t}+\frac{2m\pi}{t}, \quad 
-i \frac{\lambda_3}{t}=\frac{-4k\pi}{3t}-\frac{2m\pi}{t}.
\ee 
The characteristic polynomial of the matrix $-i(-A+P)$ in  (\ref{tildeA}) is 
\be{Carapol}
p(\lambda):=\lambda^3-(b^2+a^2+c^2+1+ab)\lambda+(a+b)(ab-c^2)+b. 
\ee 
By expressing the coefficients in terms of the desired eigenvalues 
(\ref{eigen}) \cite{Brooks}, we obtain the following two conditions for the real numbers 
$a,b,c,$ and $t>0$, 

\be{CONDIT1}
b^2+a^2+c^2+1+ab=\left( \frac{2\pi}{t}\right)^2\left( \frac{k^2}{3}+km +m^2\right), 
\ee
\be{CONDIT2}
(a+b)(ab-c^2)+b=\left( \frac{2\pi}{t}\right)^3 \left( \frac{k}{3} \left( \frac{k}{3}+m\right) \left( \frac{2k}{3}+m \right) \right). 
\ee
In order to assign the eigenvalues of $e^{At}$ corresponding to the desired operation on 
the lowest two 
energy eigenspaces,   we need to impose that the eigenvalues of $-i\begin{pmatrix} 
a & c \cr 
c & b\end{pmatrix}t$ are $i\phi_l:=i\alpha+i 2\pi  l$ and 
$i\psi_r:=i\beta+i2 \pi r$, for integers $l$ and $r$, for the desired values $\alpha$ and $\beta$.  Therefore, the symmetric matrix 
\be{tildeC}
\tilde C:=\begin{pmatrix} - a & -c \cr -c & -b \end{pmatrix}, 
\ee 
has eigenvalues $\frac{\phi_l}{t}$, $\frac{\psi_r}{t}$. The characteristic polynomial 
of the matrix $\tilde C$ in (\ref{tildeC}) is 
\be{charpol2}
p(\lambda)=\lambda^2+(a+b)\lambda+(ab-c^2).  
\ee 
Imposing that the eigenvalues are $\frac{\phi_l}{t}$, $\frac{\psi_r}{t}$, we obtain the two conditions to be added to (\ref{CONDIT1}) (\ref{CONDIT2}), 
\be{CONDIT3}
(a+b)=\frac{\phi_l+\psi_r}{t}, 
\ee
\be{CONDIT4}
ab-c^2=\frac{\phi_l \psi_r}{t^2}. 
\ee
Therefore the problem of optimal control becomes the following:
\vs

{\bf Problem 1} {\it  Given $\alpha$ and $\beta$ in 
$(-\pi, \pi]$, with both $\alpha$ and $\beta$ different from zero,\footnote{If  $(\alpha, \beta)=(0,0)$, the target final 
condition becomes the identity which is obviously reached 
in time zero. The stronger condition that both 
$\alpha$ and $\beta$ are different from zero is  used to rule out 
final conditions in the same class as $diag(e^{i\varphi},e^{-i\varphi},1)$ which are the only 
possibility if $c=0$.} find real numbers $a,b,c,t$, with $c \not=0$, $t>0$ such that conditions (\ref{CONDIT1}), (\ref{CONDIT2}), (\ref{CONDIT3}), (\ref{CONDIT4}) are verified for some integers 
$k,m,l,$ and $r$ and $t$ is the minimum such that this is possible. }

\vs

\noindent Using (\ref{CONDIT3}) and (\ref{CONDIT4}) in (\ref{CONDIT1}) and  (\ref{CONDIT2}) we obtain 
\be{DONDIT1}
\frac{(\phi_l+\psi_r)^2}{t^2}-\frac{\phi_l \psi_r}{t^2}+1=\left( \frac{2\pi}{t}\right)^2\left( \frac{k^2}{3}+km +m^2\right), 
\ee
\be{DONDIT2} 
-\frac{(\phi_l+\psi_r)(\phi_l \psi_r)}{t^3}+b=\left( \frac{2\pi}{t}\right)^3 \left( \frac{k}{3} \left( \frac{k}{3}+m\right) \left( \frac{2k}{3}+m \right) \right), 
\ee
which replace (\ref{CONDIT1}) and (\ref{CONDIT2}). Scale the time by replacing $t$ with $T:=\frac{t}{2\pi}$, and define $\hat \phi_l:=\frac{\phi_l}{2\pi}:=
\frac{\alpha}{2\pi}+l:=\hat \alpha+l$, $\hat \psi_r:=\frac{\psi_r}{2\pi}:=\frac{\beta}{2\pi}+r:=\hat \beta+r$, with $\hat \alpha:=\frac{\alpha}{2\pi}$, $\hat \beta:=\frac{\beta}{2\pi}$, both  $\in (-\frac{1}{2}, \frac{1}{2}]$. Equations (\ref{DONDIT1}), (\ref{DONDIT2}), (\ref{CONDIT3}), (\ref{CONDIT4}) therefore are written as 
\be{EONDIT1}
(\hat \phi_l+ \hat \psi_r)^2-\hat \phi_l \hat \psi_r+ T^2=\frac{k^2}{3}+km +m^2, 
\ee
\be{EONDIT2}
-(\hat \phi_l+\hat \psi_r)(\hat \phi_l \hat \psi_r)+bT^3= \frac{k}{3} \left( \frac{k}{3}+m\right) \left( \frac{2k}{3}+m \right), 
\ee
\be{EONDIT3}
(a+b)=-\frac{\hat \phi_l+\hat \psi_r}{T}, 
\ee
\be{EONDIT4}
c^2:=ab-\frac{\hat \phi_l \hat \psi_r}{T^2}>0. 
\ee
Problem 1 is therefore equivalent to, given $\hat \alpha$ and $\hat \beta$ in $(-\frac{1}{2}, \frac{1}{2}]$, finding integers $k,m,l,r$, such that $T^2>0$ in (\ref{EONDIT1}) is minimized subject to the constraint (\ref{EONDIT4}) with $b$ obtained from (\ref{EONDIT2}) and $a$ obtained 
from (\ref{EONDIT3}). We remark that, since we have assumed $c\not=0$, we cannot choose $\hat \phi_l=\hat \psi_r$. In fact, if this was the case, defining $\hat \phi_l=\hat \psi_r=\hat \gamma$, we would have, 
plugging (\ref{EONDIT3}) in (\ref{EONDIT4}), $-\left( b+ \frac{\hat \gamma}{T} \right)^2 >0$, which is a contradiction. We denote by $\hat \phi_l$ the largest of the two, i.e., we assume without loss of generality, that $\hat \phi_l > \hat \psi_r$. With this notation, the constraint of the optimization problem is equivalent  to the existence of a real $b$ satisfying (\ref{EONDIT2}), with $T^2$ given in (\ref{EONDIT1}), and satisfying  
\be{inequaonb}
-\frac{\hat \phi_l}{T}< b< -\frac{\hat \psi_r}{T}. 
\ee  
In order to see this, replace (\ref{EONDIT3}) into (\ref{EONDIT4}) to get 
\be{secondineq}
\left( b+ \frac{\hat \phi_l}{T} \right) \left( b + \frac{\hat \psi_r}{T} \right) < 0, 
\ee
which is equivalent to (\ref{inequaonb}). If we multiply (\ref{inequaonb}) by $T^3$, use (\ref{EONDIT2}) and (\ref{EONDIT1}) for $T^2$, we obtain the constraint 
\be{constr}
\hat \phi_l^3-\hat \phi_l\left(\frac{k^2}{3}+km+m^2\right) < 
\frac{k}{3} \left( \frac{k}{3} +m \right) \left( \frac{2k}{3} +m \right)  
< 
\hat \psi_r^3-\hat \psi_3 \left(\frac{k^2}{3}+km+m^2\right). 
\ee
The problem is therefore to minimize $T^2$ given in (\ref{EONDIT1}) subject to the constraint given in (\ref{constr}).  Notice that in this formulation the constraint $T^2>0$ is already included in (\ref{constr}), since $\hat \phi_l^3-\hat \phi_l\left(\frac{k^2}{3}+km+m^2\right)< \hat \psi_r^3-\hat \psi_r\left(\frac{k^2}{3}+km+m^2\right)$, with $\hat \phi_l > \hat \psi_r$ is equivalent to $\frac{k^2}{3}+km+m^2> \hat \phi_l^2+ \hat \psi_r^2+\hat \phi_l \hat \psi_r$.

Consider now the search for $k,m,l,r$ to minimize $T^2$ in (\ref{EONDIT1}) subject to (\ref{constr}).  For a given pair $l$ and $r$, $(k,m)$ and $(k,-(k+m))$ are equivalent in that they give the same value of $T^2$ and they both satisfy or do not satisfy  the constraint (\ref{constr}). Therefore we can restrict our search to values of $m \geq -\frac{k}{2}$. We define 
\be{esse}
s:=m+\frac{k}{2}\geq 0,
\ee  
which is an integer if $k$ is even and a half integer if $k$ is odd. After some algebra, write $T^2$ and the constraint in (\ref{constr}) as: 
\be{T2}
T^2:=\frac{k^2}{12}+s^2-\left( \hat \phi_l^2 + \hat \psi_r^2+ \hat \phi_l \hat \psi_r \right), 
\ee
\be{constra2}
\left( \hat \phi_l +\frac{k}{3} \right) \left(\hat \phi_l - \frac{k}{6}-s \right)\left(\hat \phi_l - \frac{k}{6}+s \right)< 0, \qquad \left( \hat \psi_r +\frac{k}{3} \right) \left(\hat \psi_r - \frac{k}{6}-s \right)\left(\hat \psi_r - \frac{k}{6}+s \right)> 0. 
\ee
So Problem 1 becomes:

\vs 
\noindent {\bf Problem 2} ({\bf Nonlinear integer optimization problem}) {\it  Given $\hat \alpha $ and $\hat \beta$ in $(\frac{1}{2}, \frac{1}{2}]$ and both different from zero,  
find $(l,r,k,s)$ with $s \geq 0$ an integer if $k$ is even and a half-integer if $k$ is odd,  to minimize $T^2$ in (\ref{T2}) with $\hat \phi_l:=\hat \alpha +l$, $\hat \psi_r:=\hat \beta+ r$, subject to the constraints in (\ref{constra2}).}
\vs

\vs 

\section{Solution to the integer optimization problem}\label{solupro}

\vs 

\begin{figure}[htb]
\centering
\includegraphics[width=0.65\textwidth, height=0.35\textheight]{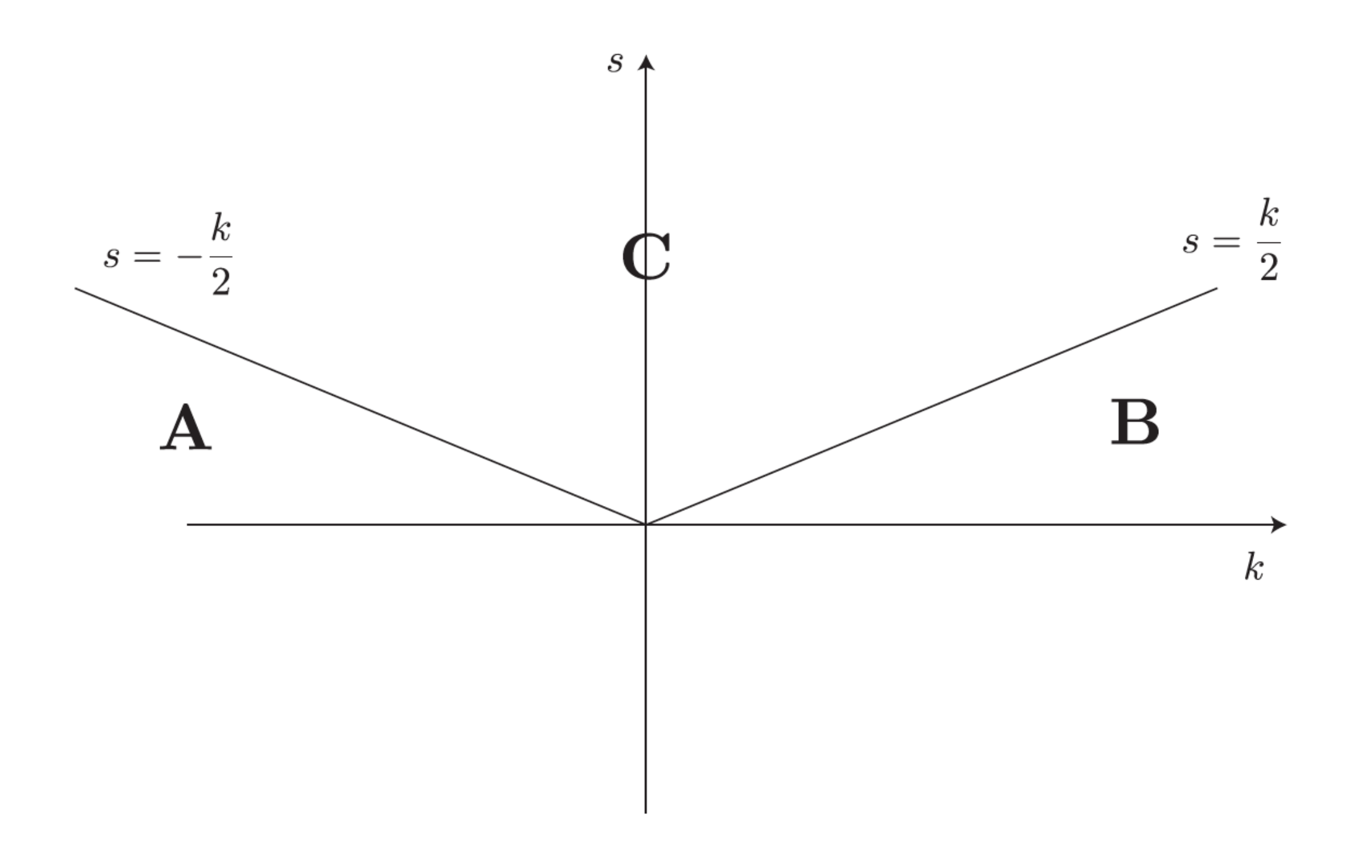}
\caption{Regions for the optimization Problem 2}.
\label{Fig111}
\end{figure}

We solve Problem $2$ as a {\it cascade} of two minimization problems. For each $(k,s)$, we 
minimize $T^2$ in (\ref{T2}) over $l$ and $r$ within the region specified by the constraint  (\ref{constra2}). Then we minimize the resulting function over $(k,s)$. Minimization over 
$(l,r)$ of $T^2$ in (\ref{T2}) for given $(k,s)$ corresponds to maximization over the same region of the function 
\be{effefunctionb}
F=F(l,r):=\hat \phi_l^2+ \hat \psi_r^2+\hat \phi_l\hat \phi_l=
\hat \alpha^2+\hat \beta^2+\hat \alpha \hat \beta +
(\hat \beta+ 2 \hat \alpha)l + (\hat \alpha+ 2 \hat \beta)r+l^2+r^2+lr. 
\ee
For given $(k,s)$, in order  to characterize the region described in constraint (\ref{constra2}), we consider all possible sign combinations which give a negative sign in the first inequality and a positive sign in the second inequality. There are four cases for each inequality. Moreover, in combining these cases we have to take into account that $\hat \phi_l > \hat \psi_r$. This reduces the possible subcases to the following three: 
\begin{enumerate}
\item {\bf A} 
\be{condiA}
\frac{k}{6} -s < \hat \psi_r < \frac{k}{6} + s < \hat \phi_l < -\frac{k}{3}; 
\ee 
\item {\bf B} 
\be{condiB}
-\frac{k}{3} < \hat \psi_r < \frac{k}{6} -s < \hat \phi_l < \frac{k}{6}+s;
\ee 
\item {\bf C} 
\be{condiC}
\frac{k}{6} -s < \hat \psi_r < - \frac{k}{3} < \hat \phi_l < \frac{k}{6}+s.
\ee 
\end{enumerate}
In order for the region described in {\bf A} to be nonempty we need $-\frac{k}{2} > s > 0$. In the case {\bf B} we need $\frac{k}{2} > s > 0$ and in the case {\bf C} we need $\frac{|k|}{2}<s$. Therefore we can solve three optimization problems according to whether $(k,s)$ are in the regions corresponding to {\bf A}, {\bf B}, or {\bf C} (cf. Figure \ref{Fig111}) and then compare the results to find the minimum. 
In fact, we shall see that there are symmetry properties so that the problems in the three regions are all equivalent in the sense that the minimum for $(k,s)$ in the region ${\bf A}$ is the same as the minimum in the region ${\bf B}$ and the minimum in the region ${\bf C}$. This will allow us to consider only one of the regions. It is a consequence of the following two (symmetry) Lemmas, whose motivation follows from the fact that (in the previous section)  the eigenvalues of $e^{(-A+P)t}$ can be permuted without changing 
the nature of the problem.

\bl{SymLemma1}
Consider the linear transformation 
\be{ktilde}
\begin{pmatrix} \tilde k \cr \tilde s \end{pmatrix}=R  \begin{pmatrix}  k \cr s  \end{pmatrix}
\ee
where $R$ is the involutory matrix $$R:=\begin{pmatrix} - \frac{1}{2} & - 3 \cr - \frac{1}{4} & \frac{1}{2} \end{pmatrix}. $$
Then $(k,s,l,r)$ is an admissible $4-$tuple with $(k,s)$ in ${\bf A}$ (${\bf C}$) if and only $(\tilde k,\tilde s,l,r)$ is an admissible $4-$tuple with $(\tilde k,\tilde s)$ in ${\bf C}$ (${\bf A}$). Moreover  $(k,s,l,r)$ and $(\tilde k,\tilde s,l,r)$ give the same value of $T^2$. 
\el 
\bpr  First notice that the transformation given by the matrix $R$  transforms  $(k, s)$ correctly into 
$(\tilde k, \tilde s)$ in the sense that if $\tilde k$ is even (odd) $\tilde s$ is an integer (half-integer).\footnote{To verify this it is enough to check all subcases: Assume first that $k$ is even and $s$ is an integer. Write $k$ as $k=-2j$. We have $\tilde k=j-3s$, $\tilde s=\frac{j}{2}+\frac{s}{2}$ from which one sees that $\tilde k$ even (odd) gives $s$ is an integer (half-integer). Analogously if $k$ is odd and $s$ is a half-integer, we can write $k=-2j+1$ and $s=\frac{1}{2}+h$, so that $\tilde k=j+h-4h-2$, $\tilde s=\frac{j+h}{2}$, from which the claim follows directly.}  Then consider $(k,s,l,r)$ satisfying equation (\ref{condiA}). Replacing $(k,s)^T$ with $(\tilde k, \tilde s)^T=R(k,s)^T$ we obtain equation (\ref{condiC}). Finally one verifies that $\frac{\tilde k^2}{12}+\tilde s^2=
\frac{k^2}{12}+s^2$, from which it follows that the value of $T^2$ is the same. 
\epr
Analogously, we have: 
\bl{SymLemma2}
Consider  the linear transformation (\ref{ktilde}) with $R$ the involutory matrix 
$$
R:=\begin{pmatrix} -\frac{1}{2} & 3 \cr \frac{1}{4} & \frac{1}{2} \end{pmatrix}
$$
Then $(k,s,l,r)$ is an admissible $4-$tuple with $(k,s)$ in ${\bf B}$ (${\bf C}$) if and only $(\tilde k,\tilde s,l,r)$ is an admissible $4-$tuple with $(\tilde k,\tilde s)$ in ${\bf C}$ (${\bf B}$). Moreover  $(k,s,l,r)$ and $(\tilde k,\tilde s,l,r)$ give the same value of $T^2$.
\el 

In view of these properties we shall reduce ourselves, without loss of generality, to optimization in the region ${\bf C}$. In the case where $\hat \beta=-\hat \alpha$, which corresponds to a matrix $\tilde X_f$ in $SU(2)$ as a final condition on the lowest two levels, we also have the following fact which allows us to reduce ourselves to the case $k\geq 0$. 
\bl{Lemmasymmetry} Assume $\hat \beta =-\hat \alpha$. Let $\{k,s,l,r\}$, and $\hat \phi_l:=\hat \alpha +l$, 
$\hat \psi_r:=-\hat \alpha+ r$ be admissible values, i.e., satisfying (\ref{condiC}), giving for the  time a value $T$. Then $\{ -k, s, -r, -l \}$ and 
$\hat \phi_{-r}:=\hat \alpha - r$, $\hat \psi_{-l}:=-\hat \alpha - l$ are also admissible values and give the same value of $T$. In particular if $(k,s)$ in region {\bf C}, to the right of the $s$-axis,  $(-k,s)$ is in  region ${\bf C}$ to the left of the $s$-axis.
\el 
\bpr If $(k,s)$ is in the region {\bf C}, taking  the negative of 
relation (\ref{condiC}) we obtain the same relation 
with $k$ replaced by $-k$ and $\hat \phi_l$ replaced by $\hat \phi_{-r}:=\alpha- r=-\hat \psi_r$ and $\hat \psi_{r}$ replaced by $\hat \psi_{-l}=-\hat \phi_l$. The time is the same. 
\epr 
We now introduce two  functions which allow us to express the constraint (\ref{condiC}) (and (\ref{condiA}), (\ref{condiB})) 
taking into account that $l$ and $r$ must be integers: The function ${\bf SI}:={\bf SI}(x)$ is the {\it smallest integer} strictly greater than $x$. The function ${\bf LI}:={\bf LI}(x)$ is the largest integer strictly smaller than $x$. The following  properties of these functions can be easily checked and will be used routinely in the following without further comment. 
\bl{SILI}
The function ${\bf SI}$ (${\bf LI}$) is nondecreasing right (left) continuous. If $p$ is an integer, then ${\bf SI}(x+p)={\bf SI}(x)+p$ (${\bf LI}(x+p)={\bf LI}(x)+p$). If $x$ is not an integer, ${\bf SI}(x)={\bf LI}(x)+1$. If $x$ is an integer,  ${\bf SI}(x)={\bf LI}(x)+2$. 
\el

\subsection{Minimization for $(k,s)$ in the region  C ($\frac{|k|}{2} < s $)}

Since $s > \frac{|k|}{2}$ and $s$ is an (half) integer for $k$ (odd) even, we have 
$s \geq \frac{|k|}{2}+1$. From equations (\ref{condiC}), we obtain 
\be{condiC1}
 \hat c:={\bf SI}\left( \frac{k}{6}-s-\hat \beta \right) \leq r \leq {\bf LI} \left( - \frac{k}{3} - \hat \beta \right) := \hat d, 
\ee
\be{condiC2}
\hat a := {\bf SI} \left( - \frac{k}{3} -\hat \alpha \right) \leq l \leq {\bf LI} \left( \frac{k}{6} + s - \hat \alpha \right):=\hat b. 
\ee
We can check, using Lemma \ref{SILI},  that generically (i.e., for $\hat \alpha$ and $\hat \beta$ in absolute value $\not= \frac{1}{3},0$) the box described in (\ref{condiC1}) (\ref{condiC2}) always contains at least one point. This is shown in Lemma \ref{Lemmanonempty} in the Appendix.

From now on we shall {\bf assume} {\mathversion{bold}${\hat \beta=-\hat \alpha}$} {\bf {and }}
{\mathversion{bold}$0<|\hat \alpha|< \frac{1}{3}$}. In particular we assume that the desired final condition $\tilde X_f$ in (\ref{Xfdes}) is (modulo a phase factor) a matrix in $SU(2)$.  Using  Lemma \ref{Lemmasymmetry}, we can restrict ourselves, 
without loss of generality, to $k \geq 0$. The function $F:=F(l,r)$ in (\ref{effefunctionb}) becomes 
\be{effefunction}
F=F(l,r)=\hat \phi_l^2+ \hat \psi_r^2+\hat \phi_l\hat \phi_l=
\hat \alpha^2+\hat \alpha (l-r) +l^2+r^2+lr.
\ee 
Proceeding to the maximization of $F=F(l,r)$ in (\ref{effefunctionb}) over the box (\ref{condiC1}) (\ref{condiC2}), the maximum  is achieved at one of the corners. The following Proposition whose proof is presented in the Appendix says which corner to use for any given pair $(k,s)$. 

\bp{SUmmarizC} Assume $\hat \alpha \in (-\frac{1}{3}, \frac{1}{3})$ and $\hat \alpha \not=0$. Write $k=6j+h$, with $h=0,1,2,...,5$, and let $(k,s)$ be in the region {\bf C} of Figure \ref{Fig111}. If $\hat \alpha >0$ the maximum of $F(l,r)$ in (\ref{effefunction}) over the admissible values of $l,r$ given in (\ref{condiC1}), (\ref{condiC2}) is given by 
$F(\hat b, \hat c)$ in all cases. 
Assume $\hat \alpha < 0$. If $h=0,3$ the maximum is still given by $F(\hat b, \hat c)$.  If  $h=2,5$, the maximum is given by $F(\hat b, \hat d)$.  If $h=1,4$ 
the maximum is at $F(\hat a, \hat c)$.
\ep 
\bpr (See the Appendix) 
\epr 

\vs 

We now obtain 
estimates for $T^2=\frac{k^2}{12}+s^2-F(l,r)$ by using 
\[ 
T^2=\frac{k^2}{12}+s^2-F(l,r)\geq \frac{k^2}{12}+s^2-\max_{l,r} F(l,r), 
\]
and for the max the values summarized in Proposition \ref{SUmmarizC} which depend on $(k,s)$.\footnote{With some abuse of notation we drop the reference to $k$ and $s$ in the $\max$.} We set once again $k=6j+h$. We have 
\be{TTT}
T^2(k,s) \geq 3j^2+jh+\frac{h^2}{12} + s^2-\max_{l,r}F(l,r).  
\ee
We separate the case $\hat \alpha > 0$ and $\hat \alpha < 0$, which require a similar analysis although the case $\hat \alpha < 0$ is a little bit more complicated because, as per 
Proposition \ref{SUmmarizC}, the maximum of $F$ is not obtained at the same corner every time. 
We present in detail the results of the analysis for $\hat \alpha > 0$ and postpone to the appendix the analysis for $\hat \alpha < 0$.

\subsubsection{Case $\hat \alpha > 0$} 
In this case, according to Proposition \ref{SUmmarizC},  the maximum of $F(l,r)$ in (\ref{TTT}) is always attained at $F(\hat b, \hat c)$. Therefore we compute  
\be{uuiu}
T^2(k,s) \geq 3j^2+jh+\frac{h^2}{12} + s^2-F(\hat b,\hat c).
\ee 
Use $\Delta_h$ equal to zero if $h$ is even and equal to $\frac{1}{2}$ if $h$ is odd. Define $\sigma_h:={\bf SI} \left( \frac{h}{6}+\Delta_h +\hat \alpha \right)$, $\lambda_h:={\bf LI} \left( \frac{h}{6}-\Delta_h - \hat \alpha \right)$. This gives, from (\ref{condiC1}) (\ref{condiC2}),  $ \hat b=j+s+\frac{1}{2}+\lambda_h$ and  $\hat c=j-s-\Delta_h+\sigma_h$. Replacing these in (\ref{uuiu}) we obtain 
\be{uuiu2}
T^2(k,s)\geq
\ee
$$ (h-3(\lambda_h+\sigma_h))j-\hat \alpha^2+\frac{h^2}{12}-\lambda_h^2 -\sigma_h^2-\lambda_h \sigma_h +(\hat \alpha+\Delta_h)(\sigma_h-\lambda_h)- \Delta_h^2+ (\sigma_h-\lambda_h-2 \Delta_h) s-2 \Delta_h\hat \alpha  -2 \hat \alpha s.
$$
Calculating the values of $\lambda_h$ and $\sigma_h$ from the definitions we obtain  
\begin{enumerate}

\item {\bf{$\bf h=0$}}
\be{CP0}
T^2\geq  -\hat \alpha^2+ (2\hat \alpha -1)+2(1- \hat \alpha)s
\geq - \hat \alpha^2 + ( 2\hat \alpha -1) + 2(1-\hat \alpha)(3j+1) \geq 1-\hat \alpha^2.   
\ee
We first used $s \geq 3j+1$ and then  $j \geq 0$. 
\item   {\bf{$\bf h=1$}}
\be{CP1}
T^2\geq -\hat \alpha^2+j+ (\hat \alpha - \frac{1}{6})+ (1-2\hat \alpha)s
\ee
$$\geq - \hat \alpha^2+(4 - 6 \hat \alpha) j -2 \hat \alpha + \frac{4}{3} \geq - \hat \alpha^2 - 2 \hat \alpha + \frac{4}{3}. 
$$
We first used $s \geq 3j+\frac{3}{2}$ and then  $j \geq 0$. 


\item   {\bf{$\bf h=2$}}
\be{CP2}
T^2\geq 
- \hat \alpha^2- j - \frac{2}{3}+(1-2 \hat \alpha)s +\hat \alpha 
\ee
$$\geq 
(-1 + 3 \hat \alpha)j+\hat \alpha+\frac{1}{3} - \hat \alpha^2 \geq 
-\hat \alpha^2+2j+\frac{4}{3}-3 \hat \alpha - 6 \hat \alpha j \geq - \hat \alpha^2+\frac{4}{3} -3 \hat \alpha. 
$$

We first used $s \geq 3j+2$ and then  $j \geq 0$. 


\item   {\bf{$\bf h=3$}}

\be{CP3}
T^2\geq 
-1-\hat \alpha^2+(2-2 \hat \alpha)s+ 2 \hat \alpha
\ee
$$\geq 
-1-\hat \alpha^2+(2-2\hat \alpha) (3j+\frac{5}{2})+ 2 \hat \alpha \geq
4-\hat \alpha^2 - 3 \hat \alpha.  
$$
We first used $s \geq 3j+ \frac{5}{2}$ and then  $j \geq 0$. 


\item   {\bf{$\bf h=4$}}

\be{CP4}
T^2\geq 3j^2+4j+\frac{4}{3}+ s^2-F(\hat b, \hat c)=
j+\frac{1}{3}-\hat \alpha^2+s-2 \hat \alpha s+ \hat \alpha
\ee
$$\geq 
(4-6 \hat \alpha) j + \frac{10}{3} -\hat \alpha^2 - 5 \hat \alpha \geq \frac{10}{3} - \hat \alpha^2 - 5 \hat \alpha.
$$

We first used  $s \geq 3j+3$ and then  $j \geq 0$. 


\item   {\bf{$\bf h=5$}}

\be{CP5}
T^2\geq  -j-\frac{7}{6}-\hat \alpha^2+(1-2 \hat \alpha)s+ \hat \alpha
\ee
$$\geq 
(2-6 \hat \alpha)j+\frac{7}{3}- \hat \alpha^2 - 6 \hat \alpha 
\geq \frac{7}{3}-6 \hat \alpha -\hat \alpha^2. 
$$

We first used $s \geq 3j+\frac{7}{2}$ and then  $j  \geq 0$. 


\vs

\end{enumerate}

\noindent All the lower bounds can be obtained with the lowest possible values of $j$ and $s$. By comparison of these lower bounds we obtain the minimum time if we assume $\hat \alpha > 0$. 

\bl{C1Theo} The minimum of $T^2$ for values of $(k,s)$ in the region {\bf C} and $\hat \alpha \in (0, \frac{1}{3})$ is given by 
$\hat T_{+,2}:=\frac{4}{3}-3 \hat \alpha- \hat \alpha^2$ if $\hat \alpha \in [\frac{1}{9}, \frac{1}{3})$. In this case 
$j=0$ and $h=2$ so that $k=6j+2=2$, $s=2$ and $l=\hat b=j+s=2$, $r=\hat c=j-s+1=-1$. 
If $\hat \alpha \in (0,\frac{1}{9})$, the minimum is  $\hat T^2_{+,0}:=1-\hat \alpha^2$.
In this case, $j=0$ so that $k=6j=0$, $s=1$, and 
$l=\hat b=j+s-1=0$, $r=\hat c=j-s+1=0$.
\el

\subsubsection{Case $\hat \alpha < 0$}


If $\hat \alpha < 0$, we follow Proposition \ref{SUmmarizC} which gives different points where the maximum of $F(l,r)$ is achieved according to the value of  $h$ in $k=6j+h$. The detailed analysis, similar to the one for the case $\hat \alpha > 0$ is presented in the Appendix. The final result corresponding to Lemma \ref{C1Theo} is as follows.

\bl{C2Theo} The minimum of $T^2$ for values of $(k,s)$ in the region {\bf C} with $k \geq 0$ and $\hat \alpha \in ( -\frac{1}{3},0)$ is given by 
$\hat T_{-,0}:=-2\hat \alpha-\hat \alpha^2$, for all values of $\hat \alpha$. 
In this case 
$j=0$ so that $k=6j=0$, $s=1$,  and $l=\hat b=j+s=1$, $r=\hat c=j-s=-1$. 
\el

\vs 

In the problem of reaching a certain matrix $\tilde X_f$ in $SU(2)$ in minimum time (cf. (\ref{Xfdes})), we can choose between  
positive and  negative $\hat \alpha$. Therefore the minimum is obtained using the minimum between the time in Lemma  \ref{C1Theo} and the one in Lemma \ref{C2Theo}. By comparison we obtain the following theorem which solves the integer optimization problem ({\bf Problem 2}) and therefore the optimal control problem for any matrix in $SU(2)$ with eigenvalues $\alpha$ and $-\alpha$, with  $0< |\alpha|=|2\pi \hat \alpha|< \frac{2\pi}{3}$. We have 

\bt{FINALE} Assume 
$0< |\hat \alpha| < \frac{1}{3}$.  If $|\hat \alpha| \leq \frac{4}{15}$ the minimum time is given by $T^2=2|\hat \alpha|- \hat \alpha^2$, with $\hat \alpha < 0$, $k=0$, $s=1$, $l=1$, $r=-1$.\footnote{It follows in these cases from (\ref{L1L2L3}) that the matrix $e^{(-A+P)t}$ is the identity.} If $|\hat \alpha| \geq \frac{4}{15}$ the minimum is attained by $T^2=\frac{4}{3}-3|\hat \alpha| - \hat \alpha^2$, for $\hat \alpha > 0$, $k=2$, $s=2$, $l=2$, $r=-1$. 
\et

\section{Conclusions}\label{concluese}
The result of Theorem \ref{FINALE} which solves the integer optimization problem allows us to find the parameters $a,b,c$ to be used in the matrix $A$ in (\ref{AandP}) and therefore the optimal control $e^{At}Pe^{-At}$ (cf. (\ref{2Bsat})). Summarizing the discussion of the above two sections, the procedure is as follows. Given the desired final condition which we assume to be of the form (\ref{Xfdes}), from the eigenvalues of $\tilde X_f$ (which we assume to be in $SU(2)$), $e^{i \alpha}$ and $e^{-i\alpha}$, we obtain $\alpha$ and $\hat{\alpha}:=\frac{\alpha}{2 \pi} \in (\frac{1}{2}, \frac{1}{2}]$. Assuming $0< |\hat \alpha| < \frac{1}{3}$, Theorem \ref{FINALE} gives the value of $T^2:=\frac{t^2}{4 \pi^2}$ (the optimal time) as well as the optimal $k$, $s:=m+\frac{k}{2}$, $l$ and $r$, and indicates whether to choose the positive or negative value for $\hat \alpha$. Using these values in (\ref{EONDIT2}) we obtain the value of $b$. Using this in (\ref{EONDIT3}) we obtain the value of $a$ and using these in (\ref{EONDIT4}) we obtain $c$. These are the parameters for the optimal control. Such a control in general will drive optimally to a state in $\hat X_f \in SU(2)$ which is only similar to the 
desired $\tilde X_f$. The corresponding matrix in $SU(3)$ will be in the same equivalence class as the desire final condition. In general, therefore there will exist a $K\in SU(2)$ such 
that $ K \hat X_f K^\dagger =\tilde X_f$. By similarity transformation of $A$ and $P$ with 
$\begin{pmatrix} 1 & 0 \cr 0 & K  \end{pmatrix}$ we obtain the optimal $A$ and $P$ matrices to be used in the optimal control (\ref{2Bsat}). 
\bex{Hadamar} Assume we want to drive in minimum time up to a scalar matrix to the final condition 
\[
H=\left(\begin{array}{ccc}
1 & 0 & 0\\
0 &\frac{1}{\sqrt{2}} & \frac{i}{\sqrt{2}}\\
0 &\frac{i}{\sqrt{2}} & \frac{1}{\sqrt{2}}\\
\end{array}\right)
\]
which performs an Hadamard-like gate \cite{NC}, 
\be{Hada}
\tilde{X}_f:=\begin{pmatrix} \frac{1}{\sqrt{2}} & \frac{i}{\sqrt{2}}\\
\frac{i}{\sqrt{2}} & \frac{1}{\sqrt{2}}\end{pmatrix},
\ee 
on the lowest two energy eigenstates. The eigenvalues of $\tilde{X}_f$ are $e^{\pm i \pi/4}$, so, using the notation of the previous section, $|\hat{\alpha}|=\frac{\pi/4}{2\pi}=\frac{1}{8}$. Using  Theorem \ref{FINALE}, we find for the minimum $T$, $T=\frac{\sqrt{15}}{8}$, $\hat \alpha=-\frac{1}{8}$ and therefore negative  $k=0$, $s=1$, $l=1$, $r=-1$. Notice that the eigenvalues of $(-A+P)t$ from (\ref{L1L2L3}) are $\lambda_1=0$, $\lambda_2=2\pi i$, $\lambda_3=-2\pi i$ so that $e^{(-A+P)t}$ is the identity. 
We also get $\hat \phi_l=\hat \alpha+l= -\frac{1}{8}+1=\frac{7}{8}$ and $\hat \psi_r=
-\hat \alpha+r= \frac{1}{8}-1=-\frac{7}{8}$. From (\ref{EONDIT2}) and (\ref{EONDIT3}) we obtain $b=0$ and $a=0$, while from (\ref{EONDIT4}) we obtain $c^2=\frac{49}{15}$. . Therefore, the equivalence class of  $H$ is reached in minimum time $t=2\pi T= \frac{\sqrt{15}\pi}{4}$ (after re-scaling of time)   using
\[
A=\left(\begin{array}{ccc}
0&0&0\\
0 &0& \pm\frac{7i}{\sqrt{15}}\\
0&\pm\frac{7i}{\sqrt{15}} & 0\\
\end{array}\right).
\]
The final condition with this control is 
\be{Finalcon555}
e^{A\frac{\sqrt{15 \pi}}{4}}
=\begin{pmatrix} 1 & 0 \cr 0 & e^{\sigma_x \frac{7 \pi}{4}}
\end{pmatrix}, 
\ee
with $\sigma_x:=\begin{pmatrix} 0 & i \cr i & 0 \end{pmatrix}$. Explicit calculation chosing the positive value for $c$ gives that 
$$
e^{\sigma_x \frac{7 \pi}{4}}=\begin{pmatrix} \cos(t) & i \sin(t) \cr i \sin(t) & \cos(t) \end{pmatrix}|_{t=\frac{7\pi}{4}}=\begin{pmatrix} \frac{1}{\sqrt{2}} & - i \frac{1}{\sqrt{2}} \cr -i \frac{1}{\sqrt{2}} & \frac{1}{\sqrt{2}}\end{pmatrix}. 
$$ 
This is not yet the desired final condition in (\ref{Hada}) but it is similar to it using the similarity transformation $K:=\texttt{diag}(i, -i)$. Therefore to obtain the optimal control we transform by similarity transformation $\texttt{diag}(1, i, -i)$ the matrix $e^{At} P e^{-At}$ (\ref{2Bsat}). Alternatively we could have used the negative value for $c$ which would have given exactly the transformation in (\ref{Hada}) without the need of an extra similarity transformation.

\begin{figure}[htb]
\centering
\includegraphics[width=0.65\textwidth, height=0.35\textheight]{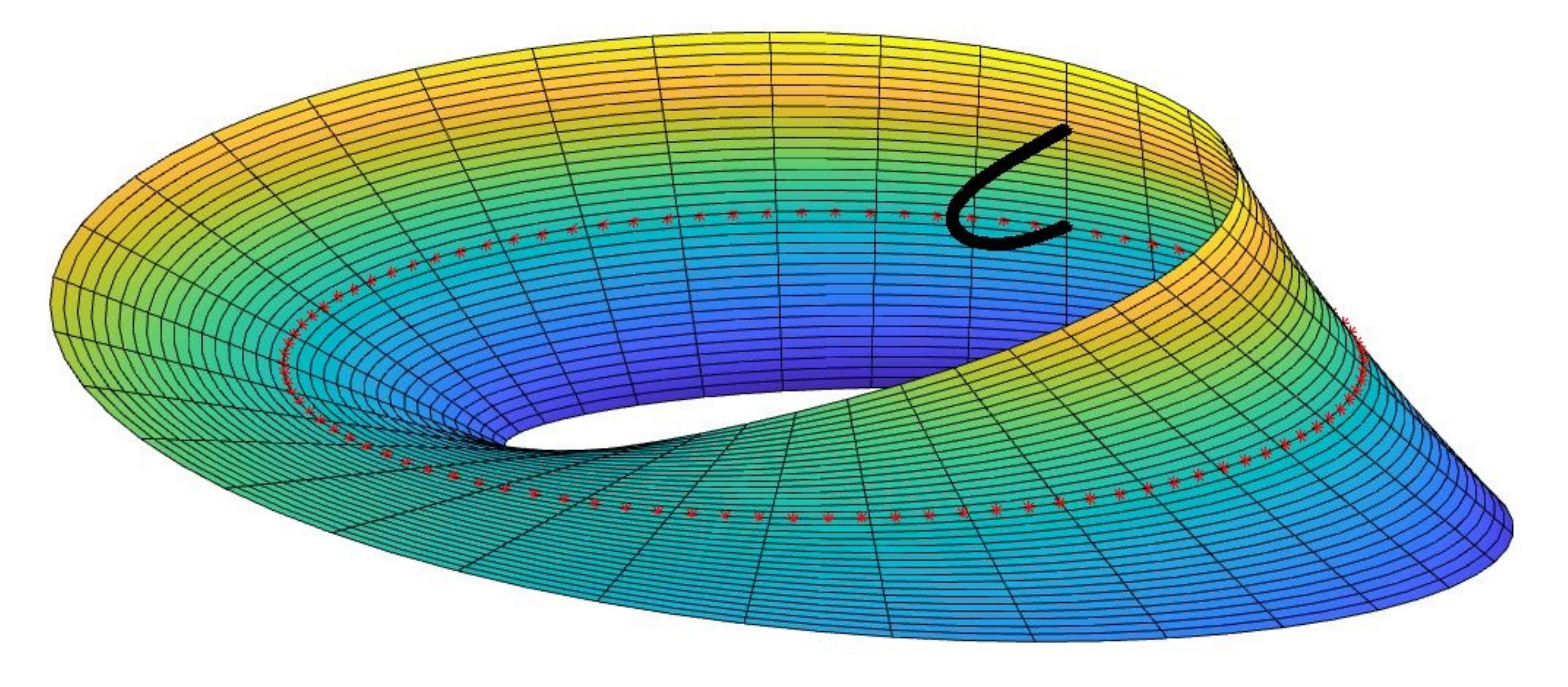}
\caption{Optimal trajectory in the quotient space}.
\label{Mobiusfig}
\end{figure}

Figure \ref{Mobiusfig} represents the optimal trajectory in the quotient space which, as we have described in section \ref{App2} can be seen as a unit base disc with at each point attached a closed (vertical)  disc and all vertical discs collapsing on segments at the border of the base disc. All the border segments form a closed M\"obius strip. In our figure we have not represented the imaginary  part of the coordinate on the vertical disc so as to be able to give a $3-D$ picture. 
For the trajectory in this example such a coordinate is identically zero. The trajectory in the quotient space for this example is represented in black. It starts from the point representing the identity and ends at the point corresponding to the final condition which is again in $E_2$ and in the same fiber as the identity. 

\eex

There are at least two reasons why we focused on final conditions in $e^{\cal K}:=S(U(n-1) \times U(1))$. From a geometrical point of view, final conditions in $S(U(n-1) \times U(1))$ belong to the {\it singular} part \cite{Bredon} of the orbit space which was described in section \ref{App2} while the regular (generic) part is the one 
corresponding to $D_1 \times D_1$ in  Theorem \ref{equiv2} and Corollary \ref{expli1}. Since the geodesics cross the regular part before reaching the final desired point, by applying results  in \cite{NOIJDCS}, \cite{Dimitry}, it follows that the geodesics lose optimality when reaching the final point. They are therefore `maximal' geodesics. They cannot be extended to a larger time interval without losing optimality. From the point of view of applications to quantum systems, it is  a common scenario for $\Lambda$-type configurations, 
to manipulate states in the subspace belonging to the lowest two energy levels. This was the problem 
which also motivated the analysis in  \cite{Ugo} where the objective  was to perform a `flip' of the states, i.e., the NOT quantum gate \cite{NC}. Our results, with a different methodology, extend the results of \cite{Ugo} 
 to a more general class of quantum gates. 

We believe that the technique used in section \ref{solupro} to solve the associated integer optimization problem, and therefore the optimal control problem, can be adapted for other ranges of the possible eigenvalues of the final conditions and therefore lead to optimal control laws for final conditions different from the ones considered in Theorem \ref{FINALE}. This, together with the maximality of the geodesics discussed above, has the potential of leading to the {\it complete optimal synthesis} for the case of $SU(3)$. Several results we have presented, such as in section \ref{App1} and section \ref{App2}, have application for general $n$ and can be used to simplify the search for the optimal control   in higher dimensional systems.  Different  
$K-P$ decompositions of $SU(n)$ could also be considered, which arise from 
 quantum systems whose energy levels are coupled by the controls in different ways. However the case of $SU(n)/S(U(n-1)\times U(1))$ considered here, is,  in some  sense, close to being the most favorable  for finding optimal trajectories. In fact,  one may use the $K-P$ structure to reduce the number of unknown parameters from $n^2-1$ to $2(n-1)$. Other $K-P$ decompositions of $SU(n)$ give rise to other reductions; however the number of parameters needed generally still grows quadratically in $n$. For example, the regular part of the quotient space of $SU(n)/e^{\cal K}$ using the decomposition $\mathfrak{su}(n)=\mathfrak{so}(n)\oplus\mathfrak{so}(n)^{\perp}$, and ${\cal K}=\mathfrak{so}(n)$,  has dimension $\frac{n^2}{2}+\frac{n}{2}-1$. 
 The knowledge of the optimal control and time for the $K-P$ case also gives lower bounds on the time needed for control for the case where a smaller number of controls is available.

\vs

\section*{Acknowledgment}
The research of Domenico D'Alessandro and Benjamin Sheller was partially supported by the NSF 
under Grant ECCS 1710558.

\appendix\newpage\markboth{Appendix}{Appendix}
\renewcommand{\thesection}{\Alph{section}}
\numberwithin{equation}{section}


\section*{APPENDIX}

\section{Completion  of the proof of Theorem \ref{equiv2}}

We are left with  proving that  $\Psi$ is i) well-defined, ii)  one-to-one, and iii) continuous with continuous inverse.

{\bf{i) $\Psi$ is well-defined.}}

Assume first that $Z, \, B \in SU(n)$ are such that $ [Z]_{\sim_{\phi}}= [B]_{\sim_{\phi}}$ and let $U\in SU(n-1)$ be the matrix that gives the equivalence (see equation (\ref{equiv1})), then we have:
\[
\Psi\left( [B]_{\sim_{\phi}}\right)= \left[ \left(\begin{array}{lllll} 
	e^{-i\phi} &  0 & 0& \cdots &0 \\
	0  & e^{i\phi} & 0& \cdots &0 \\
	0 &  0 & &&\\
	\vdots &\vdots& & I &\\
	0& 0  & & & 
	 \end{array} \right)
	  \left(\begin{array}{lc}
	1 &  0     \cdots   0 \\
	0 &   \\
	 \vdots &B \\
	  0 & \\ \end{array}\right) \right]_{\sim}=
	  \]
	  \[
	  \left[ \left(\begin{array}{lllll} 
	e^{-i\phi} &  0 & 0& \cdots &0 \\
	0  & 1 & 0& \cdots &0 \\
	0 &  0 & &&\\
	\vdots &\vdots& & I &\\
	0& 0  & & & 
	 \end{array} \right) 
	 \left(\begin{array}{lllll} 
	1&  0 & 0& \cdots &0 \\
	0  & e^{i\phi} & 0& \cdots &0 \\
	0 &  0 & &&\\
	\vdots &\vdots& & I &\\
	0& 0  & & & 
	 \end{array} \right) \left(\begin{array}{lc}
	1 &  0     \cdots   0 \\
	0 &   \\
	 \vdots &B \\
	  0 & \\ \end{array}\right) \right]_{\sim}
	  \]=
	  \[
  \left[ \left(\begin{array}{lc}
	1 &  0     \cdots   0 \\
	0 &   \\
	 \vdots &U \\
	  0 & \\ \end{array}\right) \left(\begin{array}{lllll} 
	e^{-i\phi} &  0 & 0& \cdots &0 \\
	0  & 1 & 0& \cdots &0 \\
	0 &  0 & &&\\
	\vdots &\vdots& & I &\\
	0& 0  & & & 
	 \end{array} \right) 
	 \left(\begin{array}{lllll} 
	1&  0 & 0& \cdots &0 \\
	0  & e^{i\phi} & 0& \cdots &0 \\
	0 &  0 & &&\\
	\vdots &\vdots& & I &\\
	0& 0  & & & 
	 \end{array} \right) \left(\begin{array}{lc}
	1 &  0     \cdots   0 \\
	0 &   \\
	 \vdots &B \\
	  0 & \\ \end{array}\right)
	  \left(\begin{array}{lc}
	1 &  0     \cdots   0 \\
	0 &   \\
	 \vdots &U^{\dagger} \\
	  0 & \\ \end{array}\right) \right]_{\sim}
	  \]
Since the first and the second matrices of the above expression  commute, and 
\[U\left( \begin{array}{lc}  
		e^{i\phi} & 0 \\
		0 & I \end{array} \right) BU^{\dagger} = \left( \begin{array}{lc}  
		e^{i\phi} & 0 \\
		0 & I \end{array} \right) Z,
\]
we get that
\be{equiv4}
\Psi\left( [B]_{\sim_{\phi}}\right)=\Psi\left( [Z]_{\sim_{\phi}}\right).  
\ee
So the result of applying $\Psi$ does not depend on the representative in the equivalence class $\sim_\phi$. 
Assume now that $\Psi$ acts on $D_1\times SU(n-1)/_{\sim}$. Let $Z,\, B\in SU(n-1)$  
with $[Z]_{\sim}=[B]_{\sim}$ and denote by $V\in SU(n-2)$ and $\eta\in [0,2\pi]$ the matrix and the constant such that 
if
\[
X=\left( \begin{array}{lc}  
		e^{i\eta} & 0 \\
		0 & e^{-\frac{i\eta}{n-2}}V  \end{array} \right) \in S(U(n-2)\times U(1)),
		\]
then $B=XAX^{\dagger}$. We have, for $x\in D_1$:
\[
\Psi\left(x, [B]_{\sim}\right)= 
\left[ \left(\begin{array}{cclll} 
	x &  \sqrt{1-|x|^2} & 0& \cdots &0 \\
	-\sqrt{1-|x|^2}   & x^* & 0& \cdots &0 \\
	0 &  0 & &&\\
	\vdots &\vdots& & I &\\
	0& 0  & & & 
	 \end{array} \right) \left(\begin{array}{lc}
	1 &  0     \cdots   0 \\
	0 &   \\
	 \vdots &B \\
	  0 & \\ \end{array}\right) \right]_{\sim}=
\]

{\small{
\[
\left[ \left(\begin{array}{cclll} 
	x &  \sqrt{1-|x|^2} & 0& \cdots &0 \\
	-\sqrt{1-|x|^2}   & x^* & 0& \cdots &0 \\
	0 &  0 & &&\\
	\vdots &\vdots& & I &\\
	0& 0  & & & 
	 \end{array} \right)  \left(\begin{array}{lllll} 
	1&  0 & 0& \cdots &0 \\
	0  & e^{i\eta} & 0& \cdots &0 \\
	0 &  0 & &&\\
	\vdots &\vdots& & e^{-\frac{i\eta}{n-2}}V&\\
	0& 0  & & & 
	 \end{array} \right) \left(\begin{array}{lc}
	1 &  0     \cdots   0 \\
	0 &   \\
	 \vdots &A \\
	  0 & \\ \end{array}\right)
	   \left(\begin{array}{lllll} 
	1&  0 & 0& \cdots &0 \\
	0  & e^{-i\eta} & 0& \cdots &0 \\
	0 &  0 & &&\\
	\vdots &\vdots& & e^{\frac{i\eta}{n-2}}V^{\dagger} &\\
	0& 0  & & & 
	 \end{array} \right) \right]_{\sim}.
\]}}
By writing
\[
 \left(\begin{array}{lllll} 
	1&  0 & 0& \cdots &0 \\
	0  & e^{i\eta} & 0& \cdots &0 \\
	0 &  0 & &&\\
	\vdots &\vdots& & e^{-\frac{i\eta}{n-2}}V &\\
	0& 0  & & & 
	 \end{array} \right)= \left(\begin{array}{lllll} 
	e^{i\eta} &  0 & 0& \cdots &0 \\
	0  & e^{i\eta} & 0& \cdots &0 \\
	0 &  0 & &&\\
	\vdots &\vdots& & e^{-\frac{i\eta}{n-2}}V &\\
	0& 0  & & & 
	 \end{array} \right)  \left(\begin{array}{lllll} 
	e^{-i\eta} &  0 & 0& \cdots &0 \\
	0  & 1 & 0& \cdots &0 \\
	0 &  0 & &&\\
	\vdots &\vdots& & I &\\
	0& 0  & & & 
	 \end{array} \right),   
	 \]
we get
{\small{
\[
\Psi\left(x, [B]_{\sim}\right)=
\]
\[
\left[
 \left(\begin{array}{lllll} 
	e^{i\eta} &  0 & 0& \cdots &0 \\
	0  & e^{i\eta} & 0& \cdots &0 \\
	0 &  0 & &&\\
	\vdots &\vdots& & e^{-\frac{i\eta}{n-2}}V &\\
	0& 0  & & & 
	 \end{array} \right) 
	 \left(\begin{array}{cclll} 
	x &  \sqrt{1-|x|^2} & 0& \cdots &0 \\
	-\sqrt{1-|x|^2}   & x^* & 0& \cdots &0 \\
	0 &  0 & &&\\
	\vdots &\vdots& & I &\\
	0& 0  & & & 
	 \end{array} \right) 	\left(\begin{array}{lc}
	1 &  0     \cdots   0 \\
	0 &   \\
	 \vdots &A \\
	  0 & \\ \end{array}\right)
	  \left( \begin{array}{lllll} 
	e^{-i\eta} &  0 & 0& \cdots &0 \\
	0  & e^{-i\eta} & 0& \cdots &0 \\
	0 &  0 & &&\\
	\vdots &\vdots& &e^{\frac{i\eta}{n-2}}V^{\dagger} &\\
	0& 0  & & & 
	 \end{array} \right) \right]_{\sim}\]}}
	
	And multiplying inside by $1=e^{\frac{i\eta}{n}}e^{\frac{-i\eta}{n}}$ so that the matrices on the left and right above are in $SU(n)$ yields that the above equals:
\[
\left[
 \left(\begin{array}{cclll} 
	x &  \sqrt{1-|x|^2} & 0& \cdots &0 \\
	-\sqrt{1-|x|^2}   & x^* & 0& \cdots &0 \\
	0 &  0 & &&\\
	\vdots &\vdots& & I &\\
	0& 0  & & & 
	 \end{array} \right) 	\left(\begin{array}{lc}
	1 &  0     \cdots   0 \\
	0 &   \\
	 \vdots &A \\
	  0 & \\ \end{array}\right)
	  \right]_{\sim}= \Psi\left(x, [A]_{\sim}\right)\]

{\bf{ii) $\Psi$ is one-to-one.}}

Assume, by way of contradiction, that $\Psi$ is not one-to-one. Then there exist two values
$x_1,\, x_2\in \bar{D}_1$, two matrices $A_1,\, A_2\in SU(n-1)$ and constant $\phi$ and matrix $V\in SU(n-1)$ such that:
\be{equiv12}
 \left(\begin{array}{cclll} 
	x_1 &  \sqrt{1-|x_1|^2} & 0& \cdots &0 \\
	-\sqrt{1-|x_1|^2}   & x_1^* & 0& \cdots &0 \\
	0 &  0 & &&\\
	\vdots &\vdots& & I &\\
	0& 0  & & & 
	 \end{array} \right) 	\left(\begin{array}{lc}
	1 &  0     \cdots   0 \\
	0 &   \\
	 \vdots &A_1 \\
	  0 & \\ \end{array}\right)=
	  \ee
	  \[
	\left( \begin{array}{lc}  
		e^{i\phi} & 0 \\
		0 & e^{-\frac{i\phi}{n-1}}V  \end{array} \right)   \left(\begin{array}{cclll} 
	x_2 &  \sqrt{1-|x_2|^2} & 0& \cdots &0 \\
	-\sqrt{1-|x_2|^2}   & x_2^* & 0& \cdots &0 \\
	0 &  0 & &&\\
	\vdots &\vdots& & I &\\
	0& 0  & & & 
	 \end{array} \right) 	\left(\begin{array}{lc}
	1 &  0     \cdots   0 \\
	0 &   \\
	 \vdots &A_2 \\
	  0 & \\ \end{array}\right)
	  \left( \begin{array}{lc}  
		e^{-i\phi} & 0 \\
		0 & e^{\frac{i\phi}{n-1}}V^{\dagger} \end{array} \right)
	\]
Notice that by taking $x_1,\, x_2$ in the closure of the unit disk, we can take care of both cases 
$E_n$ and $D_1\times SU(n-1)/_{\sim} $ at the same time.

Denote by $v_{jk}$ the $(j,k)$ element of the matrix $V$. By comparing the first column of the left hand side and the right hand side of the previous equality we have:
\be{equiv13}
\left( \begin{array}{c}
     x_1 \\
     -\sqrt{1-|x_1|^2} \\
     0 \\
     \vdots \\
     0 \end{array} \right)
     =
     \left( \begin{array}{c}
     x_2 \\
     -e^{-i\frac{n\phi}{n-1}}\sqrt{1-|x_2|^2}v_{11} \\
     -e^{-i\frac{n\phi}{n-1}}\sqrt{1-|x_2|^2}v_{21}  \\
     \vdots \\
    -e^{-i\frac{n\phi}{n-1}}\sqrt{1-|x_2|^2}v_{n-1,1}  \end{array} \right)
    \ee
Thus necessarily $x_1=x_2:=x$.  This means that the pre-images are either both in $E_n$ or both in 
  $D_1\times SU(n-1)/_{\sim}$.
  
  Let us first assume that they are both in $E_n$, i.e. $|x|=1$, so $x=e^{i\eta}$. It is easy to see that we can  rewrite the left hand side in (\ref{equiv12}) as:
\be{equiv14}
 \left(\begin{array}{cclll} 
	e^{i\eta} & 0 & 0& \cdots &0 \\
	0   & e^{-i\eta} & 0& \cdots &0 \\
	0 &  0 & &&\\
	\vdots &\vdots& & I &\\
	0& 0  & & & 
	 \end{array} \right) 	\left(\begin{array}{lc}
	1 &  0     \cdots   0 \\
	0 &   \\
	 \vdots &A_1 \\
	  0 & \\ \end{array}\right)=
	  \left(\begin{array}{lc}
	e^{i\eta} &  0     \cdots   0 \\
	0 &   \\
	 \vdots &M \\
	  0 & \\ \end{array}\right),  
	  \ee
with, using (\ref{equiv12}) 
\[
M= V   \left(\begin{array}{lc}
	e^{-i\eta} &  0     \cdots   0 \\
	0 &   \\
	 \vdots &I \\
	  0 & \\ \end{array}\right) A_2 V^{\dagger}.
	  \]
So we get $[A_1]_{\sim_{-\eta}}=[A_2]_{\sim_{-\eta}}$. 

Consider now the case with preimage in $D_1\times SU(n-1)/_{\sim}$, so $|x|<1$. This implies that $\sqrt{1-|x|^2}\neq 0$. Thus, by using equation 
(\ref{equiv13}), we get that $v_{j1}=0$ for $j=2,\cdots,n-1$. 
Moreover, we must have:
\[
e^{-i\frac{n\phi}{n-1}}v_{11}=1.
\]
Thus the matrix $	\left( \begin{array}{lc}  
		e^{i\phi} & 0 \\
		0 & e^{-\frac{i\phi}{n-1}}V  \end{array} \right) $ in (\ref{equiv12}) is 
	of the form:
\[
 \left(\begin{array}{cclll} 
	e^{i\phi}& 0 & 0& \cdots &0 \\
	0   & e^{-\frac{i\phi}{n-1}}v_{11} & 0& \cdots &0 \\
	0 &  0 & &&\\
	\vdots &\vdots& &\tilde{V} &\\
	0& 0  & & & 
	 \end{array} \right) = \left(\begin{array}{cclll} 
	e^{i\phi} & 0 & 0& \cdots &0 \\
	0   & e^{i\phi} & 0& \cdots &0 \\
	0 &  0 & &&\\
	\vdots &\vdots& &\tilde{V} &\\
	0& 0  & & & 
	 \end{array} \right).
	 \]
Since this matrix commutes with 
\[ \left(\begin{array}{cclll} 
	x_2 &  \sqrt{1-|x_2|^2} & 0& \cdots &0 \\
	-\sqrt{1-|x_2|^2}   & x_2^* & 0& \cdots &0 \\
	0 &  0 & &&\\
	\vdots &\vdots& & I &\\
	0& 0  & & & 
	 \end{array} \right) ,
	 \]
using equation (\ref{equiv12}), we get:
\[
A_1=\left(\begin{array}{lc}
	e^{i\phi}  &  0     \cdots   0 \\
	0 &   \\
	 \vdots &\tilde{V} \\
	  0 & \\ \end{array}\right)A_2\left(\begin{array}{lc}
	e^{-i\phi}  &  0     \cdots   0 \\
	0 &   \\
	 \vdots &\tilde{V}^{\dagger} \\
	  0 & \\ \end{array}\right),
	  \]
 for $\tilde V \in U(n-2) \times U(1)$, so $[A_1]_{\sim}=[A_2]_{\sim}$, where the equivalence  relation is now in the space of $(n-1) \times (n-1)$ matrices.

{\bf{iii)} $\Psi$ is continuous with continuous inverse}	

It suffices to prove that $\Psi$ is a continuous bijection from a compact space to a Hausdorff space, as any continuous bijection from a compact space to a Hausdorff space is a homeomorphism. Parts (ii) and (iii) above take care of the bijection part. $E_n\cup(D_1\times SU(n-1)/_{\sim})$ is compact because $E_n$ is compact as it is a fiber bundle over a compact space with compact fibers; gluing this to the boundary of $D_1$ means that $D_1\cup E_n\equiv \bar{D}_1\cup E_n$ is compact, and $SU(n-1)/_{\sim}$ is compact as it is the continuous image of the compact space $SU(n-1)$. 

$SU(n)/_{\sim}$ is Hausdorff, as it is a stratified space \cite{Bredon}; In particular, if $p,q\in SU(n)/_{\sim}$, $p\neq q$, then if $p$ and $q$ are in the same stratum, then that stratum is a manifold and hence Hausdorff, so there exist disjoint open neighborhoods of $p$ and $q$; if $p$ and $q$ are in different strata, then  we may take  each stratum as  open set.

So it remains to prove that $\Psi$ is continuous. Observe that $\Psi\circ\pi=q\circ\hat{\Psi}$, where $\pi:(U(1)\times SU(n-1))\cup (D_1\times SU(n-1))\to E_n\cup D_1\times SU(n-1)/_{\sim}$ is the quotient map which sends $(e^{i\phi},A)\in U(1)\times SU(n-1)$ to $[A]_{\phi}\in E_n$ and $(x,A)\in D_1\times SU(n-1)$ to $(x,[A]_{\sim})\in D_1\times SU(n-1)/_{\sim}$;  $q:SU(n)\to SU(n)/_{\sim}$ is the quotient map; and $\hat{\Psi}$ is defined as follows:
\be{psihat1}
\text{ if } (e^{i\phi},A)\in U(1)\times SU(n-1) \  \ \text{ then } \  \ \hat{\Psi}\left((e^{i\phi},A)\right)=
 \left(\begin{array}{lllll} 
	e^{-i\phi} &  0 & 0& \cdots &0 \\
	0  & e^{i\phi} & 0& \cdots &0 \\
	0 &  0 & &&\\
	\vdots &\vdots& & I &\\
	0& 0  & & & 
	 \end{array} \right) \left(\begin{array}{lc}
	1 &  0     \cdots   0 \\
	0 &   \\
	 \vdots & A \\
	  0 & \\ \end{array}\right)
\ee
\[
\text{ if } \left( x, A \right) \in D_1\times SU(n-1) \  \  \text{ then }  \  \  \hspace{6.5cm}
\]
\be{psihat2}
\hat{\Psi} \left( x, A \right) = 
 \left(\begin{array}{cclll} 
	x &  \sqrt{1-|x|^2} & 0& \cdots &0 \\
	-\sqrt{1-|x|^2}   & x^* & 0& \cdots &0 \\
	0 &  0 & &&\\
	\vdots &\vdots& & I &\\
	0& 0  & & & 
	 \end{array} \right) \left(\begin{array}{lc}
	1 &  0     \cdots   0 \\
	0 &   \\
	 \vdots & A \\
	  0 & \\ \end{array}\right).
\ee

Note that both $q$ and $\hat{\Psi}$ are continuous functions, and since $\pi$  is a quotient map, it is open and surjective. From this, continuity of $\Psi$ follows.

\section{Proof of Proposition \ref{SUmmarizC}}

We first fix $r=\hat c+q$, $q=0,1,2,..., \hat d- \hat c$ and maximize the function $F=F(l,r)$ over  $l \in [\hat a, \hat b]$. By comparing the values of $F(\hat a,r)$ and $F(\hat b,r)$ with $F$ in (\ref{effefunctionb}), we find that the maximum is achieved at $\hat b$ if and only if $\hat a=\hat b$ or, by writing $k:=6j+h$, $j\geq 0$, $h=0,1,2,3,4,5$, 
\be{where2Cb}
{\bf SI} \left(-\frac{h}{3}-\hat \beta \right)+ 2 {\bf SI}\left(-\frac{h}{3}-\hat \alpha \right) +q \geq \left(-\hat \beta -\frac{h}{3}\right) +2 \left(-\frac{h}{3}-\hat \alpha \right)+1.  
\ee
This condition  is independent of $s$ and $j$ and 
it is always verified when $q \geq 1$.
\bpr (Proof of  (\ref{where2Cb}))

If $\hat b > \hat a$, $r=\hat c+q$, $F(r, \hat b)\geq  F(r, \hat a)$ if and only if  
$\hat c+ q \geq -\hat \beta-2 \hat \alpha - (\hat a + \hat b)$. With the 
values of $\hat a,$ $\hat b$ and $\hat c$ in (\ref{condiC1}) (\ref{condiC2}), this latest condition becomes 
\be{whereC}
{\bf SI} \left(\frac{k}{6}-s-\hat \beta  \right) + q \geq - \hat \beta - 2 \hat \alpha - 
\left(  {\bf SI} \left(-\frac{k}{3} - \hat \alpha \right) + {\bf LI} \left( \frac{k}{6}+s -\hat \alpha \right) \right). 
\ee   
Define $\Delta_k$ to be equal to $\frac{1}{2}$ if $k$ is odd and equal to zero if $k$ is even. With this definition formula (\ref{whereC}) becomes 
\be{where1C}
{\bf SI} \left( \frac{k}{6} + \Delta_k - \hat \beta \right) + q \geq - \hat \beta - 2 \hat \alpha -\left( {\bf SI} \left( - \frac{k}{3} -\hat \alpha \right) + {\bf LI} \left( \frac{k}{6} - \Delta_k - \hat \alpha \right) \right),  
\ee
which is independent of $s$. Defining $k:=6j+h$, $h=0,1,...,5$, this becomes 
\be{where2C}
{\bf SI}\left( \frac{h}{6} + \Delta_k - \hat \beta \right) + q  \geq - \hat \beta 
- 2 \hat \alpha - 
\left( {\bf SI} \left( - \frac{h}{3} - \hat \alpha \right) + {\bf LI} \left( \frac{h}{6} - \Delta_k - \hat \alpha \right) \right), 
\ee
which is independent of $j$. To obtain (\ref{where2Cb}), write the left hand side of (\ref{where2C}) as $\Delta_k+\frac{h}{2}+ {\bf SI}\left( -\frac{h}{3} - \hat \beta \right)+q$ and the right hand side as $- \hat \beta 
- 2 \hat \alpha - 
\left( {\bf SI} \left( - \frac{h}{3} - \hat \alpha \right) -\Delta_k 
-\frac{h}{2} +{\bf LI} \left( \frac{2h}{3} - \hat \alpha \right) \right) $ 
so that inequality (\ref{where2C}) becomes $${\bf SI}\left( -\frac{h}{3} - \hat \beta \right)+q \geq - \hat \beta 
- 2 \hat \alpha - 
\left( {\bf SI} \left( - \frac{h}{3} - \hat \alpha \right)  +
{\bf LI} \left( \frac{2h}{3} - \hat \alpha \right) \right)=
$$ 
$$
- \hat \beta 
- 2 \hat \alpha - \left( {\bf SI} \left( - \frac{h}{3} - \hat \alpha \right)+h+{\bf LI} \left( - \frac{h}{3} - \hat \alpha \right)\right)= - \hat \beta 
- 2 \hat \alpha - 2{\bf SI} \left( - \frac{h}{3} - \hat \alpha \right)  +1-h,$$
where we used Lemma \ref{SILI}. This, rearranging the terms, gives    (\ref{where2Cb}). 
\epr

\vs 

By verifying  (\ref{where2Cb}) for 
the various values of $h=0,1,...,5$,\footnote{To slightly reduce the number of 
cases to check one can set $h=3p+u$, with $u=0,1,2$. This allows us to only verify 
${\bf SI}\left( \hat \alpha - \frac{u}{3} \right)+ 2 {\bf SI} \left( -\hat \alpha - \frac{u}{3} \right)+q \geq 1- \hat \alpha -u$ for $u=0,1,2$. }, we find the following 
result.
\bl{Lemmaspicchiobis}  Assume $\hat \alpha> 0$. Then the maximum of $F(r,l)$ as a function of $l$ on $[\hat a, \hat b]$ is achieved at $\hat b$ independently of the value of $r \in [\hat c, \hat d]$ and the value of $k$, since (\ref{where2Cb}) is always verified. If $\hat \alpha < 0$, the maximum is also achieved at $\hat b$ for each value of $r$ and each value of $k:=6j+h$ except for the cases $h=1$ and $h=4$. In these cases $\hat b \not=\hat a$ since $k\not=0$ and (\ref{where2Cb}) is not verified.  In these cases  the maximum is achieved at $l=\hat a$ for $r=\hat c$ and at $l=\hat b$ for all other values of $r$. 
\el

We now study $F=F(l,r)$ as a function of $r$. Assume first $\hat \alpha > 0$ or $\hat \alpha < 0$ but $h \not=1,4$. Then according to Lemma \ref{Lemmaspicchiobis} we first study $F(\hat b, r)$ for $r$ in the interval $[\hat c, \hat d]$. The maximum is achieved at $\hat c$ if and only if $F(\hat b, \hat c) \geq F(\hat b, \hat d)$, which,  using (\ref{effefunction}), is true if and only if $\hat c=\hat d$ or $\hat c+ \hat d \leq \hat \alpha - \hat b$.  The latter with the expressions for $\hat b$, $\hat c$ and $\hat d$ in (\ref{condiC1}) (\ref{condiC2}), with $\hat \beta=-\hat \alpha$ and using $k:=6j+h$ becomes: 
\be{toBverifC}
{\bf SI} \left( \frac{h}{6} +\Delta_k + \hat \alpha \right)+ {\bf LI} \left( - \frac{h}{3} +\hat \alpha \right) \leq \hat \alpha - {\bf LI} \left( \frac{h}{6}- \Delta_k -\hat \alpha \right). 
\ee
Direct verification shows that (\ref{toBverifC}) is always verified if $\hat \alpha> 0$, and therefore in these cases the maximum is at $F(\hat b, \hat c)$. It is  also verified 
for $\hat \alpha < 0$ unless $h=2$ or $h=5$. Therefore the maximum in the latter two cases is at $F(\hat b, \hat d)$ (this includes the possibility that $\hat c=\hat d$). Now consider the cases $h=1$, $h=4$  and $\hat \alpha < 0$. We have to consider two subcases: 

\noindent
1) $\hat c=\hat d$ which occurs if and only if $s=\frac{k}{2}+1$ from Lemma  \ref{Lemmanonempty}. \footnote{Recall we are assuming $k \geq 0$. Therefore the second condition of the Lemma is automatically verified.} In this case, the maximum is at $F(\hat a, \hat c)=F(\hat a, \hat d)$.

\noindent
2) $s > \frac{k}{2}+1$ and therefore $\hat d > \hat c$. Then we have to compare $F(\hat a, \hat c)$ and $F(\hat b, \hat d)$. We calculate the values of $\hat a$, $\hat b$, $\hat c$ and $\hat d$ from (\ref{condiC1}) (\ref{condiC2}) in this case for $k:=6j+1$ and $k=6j+4$ (recall  that $j\geq 0$), we obtain the following:
\begin{enumerate}
\item For $h=1$: 
\be{abcdplus}
\hat a= -2 j, \qquad \hat b=j+s-\frac{1}{2}, \qquad \hat c=j- s+\frac{1}{2}, \qquad \hat d= -2j-1. 
\ee
\item For $h=4$:
\be{abcdplus2}
\hat a= -2 j-1, \qquad \hat b=j+s, \qquad \hat c=j- s+1, \qquad \hat d= -2j-2. 
\ee

\end{enumerate}
We have
\begin{enumerate}
\item For $h=1$ 
\be{FACFBDp}
F(\hat a, \hat c)=-\hat \alpha^2+\hat \alpha (-3j+s-\frac{1}{2})+3j^2+s^2+\frac{1}{4}-s,\qquad F(\hat b, \hat d)= - \hat \alpha^2+ +\hat \alpha(3j+s+\frac{1}{2})+3j^2+3j+s^2+\frac{7}{4}-2s. 
\ee  
By comparison we have that $F(\hat a, \hat c) \geq  F(\hat b, \hat d)$ if and only if 
$ s \geq 3j+\frac{3}{2}+\hat \alpha (6j+1)$, which is indeed true since (using $s \geq \frac{k}{2}+1$)
$s \geq 3j+ \frac{3}{2}$ and $\hat \alpha<0$. 

\item For $h=4$ 
\be{FACFBDm}
F(\hat a, \hat c)=-\hat \alpha^2+\hat \alpha(-3j+s-2)+ 3 j^2+3j+1+s^2-s       \qquad          F(\hat b, \hat d)= -\hat \alpha^2+\hat \alpha(3j+s+2)+3j^2+s^2+6j+4-2s
\ee  
By comparison we have that $F(\hat a, \hat c) \geq  F(\hat b, \hat d)$ if and only if 
$ s \geq 3j+3+\hat \alpha (6j+4)$, which is indeed true since (using $s \geq \frac{k}{2}+1$)
$s \geq 3j+ 3 $ and $\hat \alpha<0$. Therefore in this case also we have that the maximum is achieved at $F(\hat a, \hat c)$. 
\end{enumerate}

\section{The box (\ref{condiC1}) (\ref{condiC2}) is not empty}

\bl{Lemmanonempty} Assume $\hat \alpha$ and $\hat \beta$ $\not=0, \pm \frac{1}{3}$.  
In (\ref{condiC1}) $\hat d > \hat c$ always unless $s=\frac{|k|}{2}+1$ and $k \geq 0$, in which case $\hat c=\hat d$.  
In (\ref{condiC2})  $\hat b > \hat a$ always unless $s=\frac{|k|}{2}+1$ and $k \leq 0$, in which case $\hat a=\hat b$. 
\el
\bpr
We prove the first statement since the proof for the second statement is similar. Using $s \geq \frac{|k|}{2}+1$, we have for $\hat c$ in (\ref{condiC1})  
$\hat c:={\bf SI} \left( \frac{k}{6} -s -\hat \beta \right) \leq 
{\bf SI} \left( \frac{k}{6} - \frac{|k|}{2} -1 -\hat \beta \right) = -1 + 
{\bf SI} \left( \frac{k}{6} - \frac{|k|}{2}  -\hat \beta \right)$, 
where the equality holds if and only if $s=\frac{|k|}{2}+1$.  If $k<0$ the last term is 
$-1 + {\bf SI}(\frac{2}{3} k -\hat \beta)=-1+k+{\bf{SI}}(-\frac{k}{3}-\hat \beta)=k+{\bf{LI}}(-\frac{k}{3}-\hat \beta)<{\bf{LI}}(-\frac{k}{3}-\hat \beta):=\hat d$. However if $k \geq 0$, 
the last term is $-1+ {\bf SI}(-\frac{k}{3} -\hat \beta)=
{\bf LI}(-\frac{k}{3} -\hat \beta):=\hat d$. 
\epr

\section{Case $\hat \alpha < 0$ (Proof of Lemma \ref{C2Theo})}

Set $k=6j+h$, with $h=0,1,2,3,4,5$ and use (\ref{condiC1}) (\ref{condiC2}). 

\begin{enumerate}

\item {\bf{$\bf h=0$}}, ($ \hat b=j+s$, $\hat c=j-s$, $j \geq 0$, $s \geq 3j +1$)

\be{CN0}
T^2\geq 3j^2+s^2-F(\hat b, \hat c)=-\hat \alpha^2 -2\hat \alpha \geq 
- \hat \alpha^2 - 2 \hat \alpha j(3j+1) \geq - \hat \alpha^2 - 2 \hat \alpha. 
\ee
Here we used first $s \geq 3j+1$ and then  $j \geq 0$. The minimum 
$\hat T^2_{-,0}:=-\hat \alpha^2 - 2 \hat \alpha$ is achieved for $j=0$ and $s=1$.

\item   {\bf{$\bf h=1$}}, ($ \hat a=-2j$, $\hat c=j-s+\frac{1}{2}$) 

\be{cN1A}
T^2\geq 3j^2+j+\frac{1}{12}+ s^2-F(\hat a, \hat c)=- \hat \alpha^2 - \frac{1}{6} + (1-\hat \alpha)s + (1+3 \hat \alpha)j+\frac{\hat \alpha}{2}
\ee
$$
\geq - \hat \alpha^2-\hat \alpha +\frac{4}{3}+3 j \geq - \hat \alpha^2-\hat \alpha +\frac{4}{3}. 
$$
Here we used first $s \geq 3j+1$ and then  $j \geq 0$. The minimum is 
$\hat T^2_{-,1}:=\frac{4}{3}- \hat \alpha-\hat \alpha^2$ is achieved for $j=0$ and $s=3j+\frac{3}{2}$.

\item   {\bf{$\bf h=2$}}, ($ \hat b=j+s$, $\hat d=-2j-1$)

\be{N2}
T^2\geq 3j^2+2j+\frac{1}{3}+ s^2-F(\hat b, \hat d)=
-(1+3 \hat \alpha) j - \frac{2}{3} - \hat \alpha - \hat \alpha^2 +(1-\hat \alpha)s 
\ee
$$\geq (2-6\hat \alpha) j + \frac{4}{3}-3 \hat \alpha - \hat \alpha^2 \geq 
\frac{4}{3}-3 \hat \alpha - \hat \alpha^2. 
$$

Here we first used we used $s \geq 3j+2$ and then  we used $j \geq 0$. The minimum $\hat T^2_{-,2}:=\frac{4}{3}- 3 \hat \alpha-\hat \alpha^2$ is achieved for $j=0$ and $s=3j+2$.

\item   {\bf{$\bf h=3$}}, ($ \hat b=j+s+\frac{1}{2}$, 
$\hat c=j-s+\frac{1}{2}$)

\be{CN3}
T^2\geq 3j^2+3j+\frac{3}{4}+ s^2-F(\hat b, \hat c)=- \hat \alpha^2-2  \hat \alpha s 
\ee
$$\geq -\hat \alpha^2 -6 \hat \alpha j -5  \hat \alpha \geq -\hat \alpha^2- 
5 \hat \alpha
$$

We first used $s \geq 3j +\frac{5}{2}$ and then $j \geq 0$. The minimum 
$\hat T^2_{-,3}:=-\hat \alpha^2- 
5 \hat \alpha$ is achieved for $j=0$ and $s=\frac{5}{2}$.


\item   {\bf{$\bf h=4$}}, ($ \hat a=-2j-1$, $\hat c=j-s+1$)

\be{CN4}
T^2\geq 3j^2+4j+\frac{4}{3}+ s^2-F(\hat a, \hat c)= 
(1+3 \hat \alpha)j + (1-\hat \alpha)s+2 \hat \alpha +\frac{1}{3}-\hat \alpha^2 
\ee
$$\geq  4j-\hat \alpha +\frac{10}{3} -\hat \alpha^2  \geq \frac{10}{3} -\hat \alpha -\hat \alpha^2.  
$$

We first used  $s \geq  3j + 3$. and then $j \geq 0$. The minimum $\hat T^2_{-,4}:=\frac{10}{3}- \hat \alpha -\hat \alpha^2$ is achieved for $j=0$ and $s=3$.

\item   {\bf{$\bf h=5$}}, ($ \hat b=j+s+\frac{1}{2}$, $\hat d=-2j-2$) 

\be{CN5}
T^2\geq 3j^2+5j+\frac{25}{12}+ s^2-F(\hat b, \hat d)=-\frac{7}{6}+ (3 -3 \hat \alpha)j-\hat \alpha^2+(1-\hat \alpha)s-\frac{5}{2}\hat \alpha.
\ee
$$\geq 
(6-6 \hat \alpha)j -\hat \alpha^2 - 6 \hat \alpha +\frac{7}{3}
\geq \frac{7}{3}-\hat \alpha^2 - 6 \hat \alpha
$$

We first used $s \geq 3j+\frac{7}{2}$ and then $j \geq 0$. The minimum $\hat T^2_{-,5 }:=\frac{7}{3}-6 \hat \alpha -\hat \alpha^2 $ is achieved for $j=0$ and $s=\frac{7}{2}$.

\end{enumerate}

\noindent By comparison of the functions $\hat T^2_{-,w}:=T^2_{-,w}(\hat \alpha)$, for $w=0,1,...,5$, 
we obtain Lemma \ref{C2Theo}.

\end{document}